\documentclass[journal=jctc,manuscript=article]{achemso}

\usepackage[T1]{fontenc}
\usepackage[utf8]{inputenc}
\usepackage{lmodern}
\usepackage{microtype}
\usepackage[hidelinks]{hyperref}

\usepackage{amsmath,amssymb,amsfonts}
\usepackage{bm}
\usepackage{braket}
\usepackage[version=4]{mhchem}
\usepackage{siunitx}
\usepackage{graphicx}
\usepackage{booktabs}
\usepackage{array}
\usepackage{multirow}
\usepackage{longtable}

\usepackage{algorithm}
\usepackage{algpseudocode}

\usepackage[nameinlink,noabbrev]{cleveref}

\SectionNumbersOn

\title{A Deterministic Framework for Neural Network Quantum States in Quantum Chemistry}

\author{Zheng Che}
\affiliation{
  Hefei National Research Center for Physical Sciences at the Microscale,
  University of Science and Technology of China, Hefei 230026, China
}
\email{wsmxcz@gmail.com}

\begin{tocentry}
  \centering
  \includegraphics[width=8.25cm, height=4.45cm, keepaspectratio]{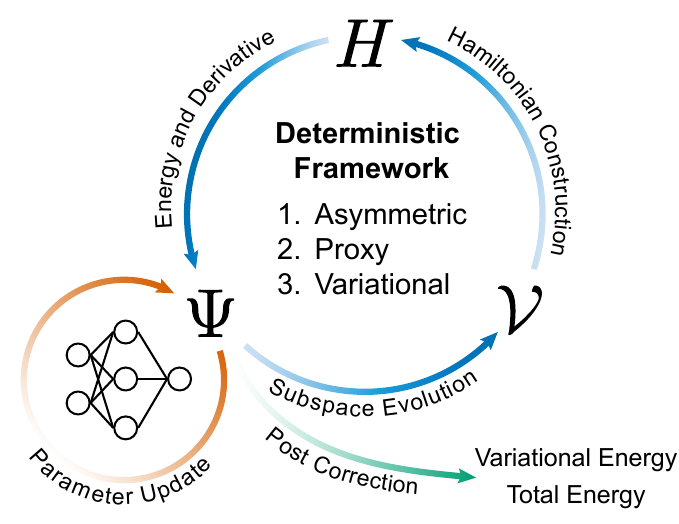}
\end{tocentry}

\begin{document}

\begin{abstract}
We present a deterministic optimization framework for Neural Network Quantum States (NQS) designed to bypass the sampling variance and slow mixing issues inherent in stochastic optimization. By projecting a neural backflow ansatz onto dynamically evolving configuration subspaces and applying a post-hoc second-order perturbative correction, our method provides a systematic route for optimizing the selected variational component of the wavefunction and estimating residual correlation through a post-hoc perturbative correction. The implementation utilizes a hybrid CPU-GPU architecture that shows empirical sub-linear wall-time scaling with respect to the subspace size over the tested range, enabling the calculation of strongly correlated systems, such as the chromium dimer, within Hilbert spaces of $10^{23}$ configurations. Benchmarks on molecular bond dissociations demonstrate that this deterministic approach yields stable convergence and accuracies comparable to selected reference methods in the tested systems.
\end{abstract}

\section{Introduction}

Solving the electronic Schrödinger equation remains a central challenge in \textit{ab initio} quantum chemistry. While Full Configuration Interaction (FCI) is exact, its exponential scaling necessitates approximations\cite{sherrill1999configuration, helgaker2013molecular, szabo2012modern}. Selected Configuration Interaction (sCI)\cite{huron1973iterative, holmes2016heat, tubman2020modern, zhang2020iterative} iteratively constructs variational subspaces of dominant configurations, whereas Coupled Cluster (CC)\cite{raghavachari1989fifth, bartlett2007coupled} and Matrix Product States (MPS)\cite{white1999ab, chan2011density} employ compact non-linear ansätze. Recently, Neural Network Quantum States (NQS) have emerged as a promising alternative, leveraging deep neural networks to parameterize many-body wavefunctions with polynomial scaling\cite{carleo2017solving, deng2017quantum, choo2020fermionic, sharir2022neural, medvidovic2024neural, lange2024architectures, hermann2023ab}.

Optimizing NQS in second-quantized chemistry, however, is hindered by sampling difficulties. Unlike continuous real-space formulations\cite{hermann2020deep, pfau2020ab, hermann2023ab}, molecular wavefunctions in discrete Fock space exhibit sparse and multimodal probability distributions. Consequently, conventional Markov Chain Monte Carlo (MCMC) sampling suffers from slow mixing and high variance\cite{choo2020fermionic, li2023nonstochastic, liu2024nnbf, shang2025solving}, often missing relevant tail configurations. Although autoregressive architectures permit direct sampling, they impose rigid orderings that structurally constrain the ansatz. Ultimately, the high variance of stochastic gradient estimators in large, sparse Hilbert spaces obscures the intrinsic representational capacity of the neural network\cite{sharir2020deep, barrett2022autoregressive, shang2025solving, luo2023gauge, ibarra2025autoregressive}.

To circumvent these sampling limitations, recent deterministic or semi-deterministic strategies\cite{li2023nonstochastic, li2024improved, liu2024nnbf, liu2025efficient} replace stochastic estimation with exact summation over selected target sets. Despite successes in small systems, different formulations of the deterministic energy objective and its gradient exhibit varying theoretical and computational properties. Directly adapting the VMC energy functional with subspace truncation often introduces a mathematical inconsistency between the objective and its gradient (i.e., ``Asymmetric'' estimators). Furthermore, while restricting optimization to a truncated subspace has been explored as a practical technique to reduce the $O(N^4)$ computational overhead\cite{li2024improved}, a comparative analysis of these disparate optimization modes is necessary to understand their convergence behaviors.

In this work, we present a deterministic framework for NQS optimization that evaluates these diverse objective formulations. Projecting the ansatz onto a dynamically evolving configuration set reformulates the Schrödinger equation as a non-linear optimization problem. By comparing three distinct optimization modes---Asymmetric, Proxy, and Variational---we analyze their theoretical consistency, gradient properties, and optimization dynamics. Based on this analysis, we find that a protocol combining strictly Variational optimization within a compact subspace and a post-hoc second-order Epstein--Nesbet perturbation (PT2) correction provides a stable and computationally practical strategy in the present implementation. This approach decouples the optimization of the selected variational component from the post-hoc estimation of residual perturbative contributions, reducing training complexity while eliminating theoretical inconsistencies.

To scale this framework, we address the computational disparity between the combinatorial logic of sparse Hamiltonian generation and the high arithmetic intensity of neural tensor contractions. We develop a hybrid architecture integrating a combinatorial kernel with a differentiable core. Combined with heat-bath screening to prune the Hamiltonian graph, this architecture shows empirical sub-linear wall-time scaling with respect to subspace size over the tested range.

We validate the framework on the bond dissociation of \ce{H2O} and \ce{N2}, and the strongly correlated chromium dimer (\ce{Cr2}). The Variational+PT2 protocol yields energies comparable to selected multi-reference benchmarks in the tested systems in Hilbert spaces exceeding $10^{23}$ dimensions.

The paper is organized as follows: Section 2 establishes the theoretical foundation and optimization modes. Section 3 details the hybrid algorithmic architecture and screening techniques. Section 4 presents numerical benchmarks, and Section 5 concludes the work.

\section{Theory}

We present the deterministic optimization framework for the time-independent electronic Schrödinger equation. We first define the Fock space notation and the geometry of the configuration sets (\cref{notation}), followed by the formulation of the Neural Network Quantum States (\cref{subsec:ansatz}). Subsequently, we define the deterministic energy objectives (\cref{subsec:modes}) and outline the self-consistent optimization procedure (\cref{subsec:evolution}).

\subsection{Framework and Notation}
\label{notation}

We consider the fermionic Fock space restricted to the sector with fixed total electron number $N_e$ and spin projection $S_z$. The space is spanned by the orthonormal Fock basis $\mathcal{B} = \{\ket{\mathbf{x}}\}$, where each basis vector is identified by an occupation bitstring $\mathbf{x} \in \{0,1\}^{N_o}$. To avoid ambiguity with the configurational structure of the ansatz described in \cref{subsec:ansatz}, we refer to the basis elements $\ket{\mathbf{x}}$ strictly as \textit{configurations}. The many-body variational wavefunction is parameterized as a linear expansion $\ket{\Psi_\theta} = \sum_{\mathbf{x} \in \mathcal{B}} \psi_\theta(\mathbf{x}) \ket{\mathbf{x}}$, where $\psi_\theta(\mathbf{x})$ denotes the real-valued amplitude computed by the neural network with trainable parameters $\theta$.

The optimization proceeds iteratively. At each outer iteration $k$, the wavefunction is supported on a variational set $\mathcal{V}_k \subset \mathcal{B}$. We define the connected set $\mathcal{C}_k$ as the subset of all configurations coupled to $\mathcal{V}_k$ via the Hamiltonian $\hat{H}$:
\begin{equation}
    \mathcal{C}_k = \{ \ket{\mathbf{x}'} \in \mathcal{B} \mid \exists \ket{\mathbf{x}} \in \mathcal{V}_k,\; \bra{\mathbf{x}'} \hat{H} \ket{\mathbf{x}} \neq 0 \}.
    \label{eq:connected_set}
\end{equation}
Equation~\ref{eq:connected_set} defines the Hamiltonian-connected search space used for deterministic subspace expansion.
The external configurations are isolated in the perturbative set $\mathcal{P}_k = \mathcal{C}_k \setminus \mathcal{V}_k$. The union $\mathcal{T}_k = \mathcal{V}_k \cup \mathcal{P}_k$ forms the target set for deterministic evaluation.

During the inner optimization loop, $\mathcal{T}_k$ remains fixed. The parameters $\theta$ are updated to minimize an energy objective defined over this target set. Upon convergence, a scoring function $w(\mathbf{x}) = |\psi_\theta(\mathbf{x})|$ is evaluated for all $\ket{\mathbf{x}} \in \mathcal{T}_k$ to select the updated variational set $\mathcal{V}_{k+1}$ for the subsequent iteration. Table~\ref{tab:symbols} summarizes the notation used throughout this work.

\begin{table}[htbp]
  \centering
  \caption{Nomenclature and symbols used in this work.}
  \label{tab:symbols}
  \renewcommand{\arraystretch}{1.2}
  \setlength{\tabcolsep}{6pt}
  \begin{tabular}{@{}ll@{}}
    \toprule
    \textbf{Symbol} & \textbf{Description} \\
    \midrule
    \multicolumn{2}{l}{\textit{System and Wavefunction}} \\
    $N_o, N_e$ & Number of spin-orbitals and electrons. \\
    $\mathbf{x} \in \{0,1\}^{N_o}$ & Occupation bitstring representing a configuration. \\
    $\ket{\mathbf{x}}$ & Orthonormal Fock basis vector. \\
    $\theta$ & Trainable parameters of the ansatz. \\
    $\psi_\theta(\mathbf{x})$ & Amplitude mapping $\mathbf{x} \mapsto \mathbb{R}$. \\
    $\ket{\Psi_\theta}$ & Global variational wavefunction. \\
    \midrule
    \multicolumn{2}{l}{\textit{Subset}} \\
    $\mathcal{V}_k$ & Variational set at iteration $k$. \\
    $\mathcal{C}_k$ & Connected set coupled to $\mathcal{V}_k$. \\
    $\mathcal{P}_k$ & Perturbative set, defined as $\mathcal{C}_k \setminus \mathcal{V}_k$. \\
    $\mathcal{T}_k$ & Target set for deterministic evaluation, $\mathcal{V}_k \cup \mathcal{P}_k$. \\
    \midrule
    \multicolumn{2}{l}{\textit{Hamiltonians and Objectives}} \\
    $\mathbf{H}_{\mathcal{XY}}$ & Hamiltonian matrix block with rows in $\mathcal{X}$ and columns in $\mathcal{Y}$. \\
    $\tilde{\mathbf{H}}_k$ & Proxy Hamiltonian defined on $\mathcal{T}_k$. \\
    $E_{\mathrm{var/asym/proxy}}$ & Variational, Asymmetric, and Proxy energy objectives. \\
    $\Delta E_{\mathrm{PT2}}$ & Second-order perturbative correction. \\
    \bottomrule
  \end{tabular}
\end{table}

\subsection{Neural Network Quantum States}
\label{subsec:ansatz}

We parameterize the wavefunction amplitude $\psi_\theta(\mathbf{x})$ using a NQS. A defining characteristic of NQS in fermionic systems is the enforcement of antisymmetry. Current architectures can be categorized into three classes based on their treatment of the sign structure:

\begin{enumerate}
    \item \textbf{Unconstrained Architectures}: General-purpose function approximators, such as RBMs or MLPs, map configurations directly to amplitudes, $\psi_\theta(\mathbf{x}) \approx \text{NN}(\mathbf{x})$\cite{choo2020fermionic, barrett2022autoregressive, li2023nonstochastic, shang2025solving}. Lacking an intrinsic antisymmetry mechanism, these models typically require complex-valued outputs or separate amplitude-phase networks to learn the sign structure. The resulting optimization landscape is highly non-convex, often necessitating pre-training or geometry-aware optimizers like Stochastic Reconfiguration\cite{wang2024variational, sorella1998green, park2020geometry}.

    \item \textbf{Slater--Jastrow Architectures}: These ansätze decompose the amplitude into a multiplicative Jastrow factor $J_\theta(\mathbf{x})$ and a fixed reference determinant: $\psi_\theta(\mathbf{x}) = e^{J_\theta(\mathbf{x})} \det(\mathbf{\Phi}_{\text{HF}}[\mathbf{x}])$. While this strictly enforces the sign structure of the reference state (e.g., Hartree--Fock), the fixed nodal surface limits accuracy in strongly correlated regimes where the ground-state sign structure deviates from the mean-field reference\cite{nomura2017restricted, humeniuk2023autoregressive}.

    \item \textbf{Neural Network Backflow Architectures}: Backflow transformations generalize the Slater determinant by making the single-particle orbitals dependent on the many-body configuration $\mathbf{x}$. By learning a configuration-dependent orbital basis $\tilde{\phi}_i(\mathbf{x})$, these architectures allow continuous deformation of the sign structure while preserving the determinant antisymmetry\cite{luo2019backflow, liu2024nnbf}.
\end{enumerate}

In this work, we employ a single-determinant neural network backflow ansatz. Assuming a time-reversal symmetric Hamiltonian, the ground-state wavefunction can be taken as real-valued; thus, we utilize a real-valued neural network.

The ansatz is constructed from a static reference orbital matrix $\mathbf{\Phi}^0 \in \mathbb{R}^{N_o \times N_e}$ associated with the Hartree--Fock reference. The backflow transformation introduces configuration dependence via a neural network $\Delta\mathbf{\Phi}_\theta: \{0,1\}^{N_o} \to \mathbb{R}^{N_o \times N_e}$. The effective orbital matrix $\mathbf{\Phi}_\theta(\mathbf{x})$ is defined as:
\begin{equation}
    \mathbf{\Phi}_\theta(\mathbf{x}) = \mathbf{\Phi}^0 + \Delta\mathbf{\Phi}_\theta(\mathbf{x}),
    \label{eq:effective_orbital_matrix}
\end{equation}
where $\Delta\mathbf{\Phi}_\theta(\mathbf{x})$ is a non-linear additive correction. Eq.~\ref{eq:effective_orbital_matrix} incorporates configuration-dependent many-body correlations into the single-particle orbital representation.

For a configuration $\mathbf{x}$, let $\mathcal{I}(\mathbf{x}) = \{i \mid x_i = 1\}$ denote the set of occupied orbital indices. The occupied submatrix $\mathbf{A}_\theta(\mathbf{x}) \in \mathbb{R}^{N_e \times N_e}$ is constructed by selecting the rows of $\mathbf{\Phi}_\theta(\mathbf{x})$ indexed by $\mathcal{I}(\mathbf{x})$:
\begin{equation}
    [\mathbf{A}_\theta(\mathbf{x})]_{jk} = 
    [\mathbf{\Phi}_\theta(\mathbf{x})]_{i_j, k}, 
    \quad \text{where } i_j \in \mathcal{I}(\mathbf{x}).
    \label{eq:occupied_submatrix}
\end{equation}
The wavefunction amplitude is the determinant of this submatrix:
\begin{equation}
    \psi_\theta(\mathbf{x}) = \det \left( \mathbf{A}_\theta(\mathbf{x}) \right).
    \label{eq:backflow_amplitude}
\end{equation}
Eqs.~\ref{eq:occupied_submatrix} and \ref{eq:backflow_amplitude} define the determinant-valued neural backflow amplitude used in this work.

The network is implemented as a residual mapping. Its output layer is initialized with a small scale so that $\Delta\mathbf{\Phi}_\theta(\mathbf{x}) \approx \mathbf{0}$ at the start of optimization. Consequently, the initial ansatz remains close to the Hartree--Fock reference while allowing the orbital representation to relax during training.

\subsection{Deterministic Energy Objectives}
\label{subsec:modes}

We formulate the optimization problem by replacing stochastic expectation values with deterministic summations. The energy objective is defined as a functional $E(\theta; \Omega, \hat{H}_{\text{eff}})$, specified by a normalization domain $\Omega$ and an effective Hamiltonian $\hat{H}_{\text{eff}}$. The general form is given by the Rayleigh quotient:
\begin{equation}
    E(\theta; \Omega, \hat{H}_{\text{eff}}) =
    \frac{\bra{\Psi_\Omega} \hat{H}_{\text{eff}} \ket{\Psi_{\mathcal{T}}}}{\braket{\Psi_\Omega|\Psi_\Omega}}
    =
    \frac{\sum_{\mathbf{x} \in \Omega} \psi_\theta(\mathbf{x}) \bra{\mathbf{x}} \hat{H}_{\text{eff}} \ket{\Psi_{\mathcal{T}}}}{\sum_{\mathbf{x} \in \Omega} |\psi_\theta(\mathbf{x})|^2},
    \label{eq:general_rayleigh_objective}
\end{equation}
where $\Psi_{\mathcal{T}}$ denotes the wavefunction restricted to the target set $\mathcal{T}_k$. The three optimization modes analyzed below are obtained from Eq.~\ref{eq:general_rayleigh_objective} by choosing different normalization domains and effective Hamiltonians.

\subsubsection{Asymmetric Mode}

This mode corresponds to deterministic schemes used in recent non-stochastic or semi-stochastic approaches~\cite{li2023nonstochastic, liu2024nnbf, kan2025nnqs}. The normalization is restricted to the variational set ($\Omega = \mathcal{V}_k$), while the Hamiltonian expectation value includes contributions from the full target set ($\hat{H}_{\text{eff}} = \hat{P}_{\mathcal{T}} \hat{H} \hat{P}_{\mathcal{T}}$). The energy is computed as the expectation value of the local energy $E_{\mathrm{loc}}(\mathbf{x})$:
\begin{equation}
    E_{\mathrm{asym}}(\theta) = \sum_{\mathbf{x} \in \mathcal{V}_k} p_{\mathcal{V}}(\mathbf{x}) E_{\mathrm{loc}}(\mathbf{x}), \quad
    E_{\mathrm{loc}}(\mathbf{x}) = \sum_{\mathbf{x}' \in \mathcal{T}_k} H_{\mathbf{x}\mathbf{x}'} \frac{\psi_\theta(\mathbf{x}')}{\psi_\theta(\mathbf{x})},
    \label{eq:asymmetric_energy}
\end{equation}
with $p_{\mathcal{V}}(\mathbf{x}) = |\psi_\theta(\mathbf{x})|^2 / \sum_{\mathbf{y} \in \mathcal{V}_k} |\psi_\theta(\mathbf{y})|^2$.
A gradient estimator analogous to the VMC log-derivative trick is typically employed:
\begin{equation}
    \mathbf{g}_{\mathrm{asym}} = 2 \sum_{\mathbf{x} \in \mathcal{V}_k} p_{\mathcal{V}}(\mathbf{x}) \left( E_{\mathrm{loc}}(\mathbf{x}) - E_{\mathrm{asym}} \right) \nabla_\theta \ln |\psi_\theta(\mathbf{x})|.
    \label{eq:asymmetric_gradient}
\end{equation}
Note that $\mathbf{g}_{\mathrm{asym}}$ in Eq.~\ref{eq:asymmetric_gradient} is not the exact gradient of $E_{\mathrm{asym}}(\theta)$ in Eq.~\ref{eq:asymmetric_energy}, as it neglects terms arising from the dependence of $E_{\mathrm{loc}}(\mathbf{x})$ on the parameters via amplitudes in $\mathcal{P}_k$. Consequently, the resulting vector field is non-conservative. The derivation of the exact full gradient is provided in the Supporting Information.

\subsubsection{Proxy Mode}

The Proxy mode restores consistency between the energy functional and its gradient by extending the normalization domain to the full target set ($\Omega = \mathcal{T}_k$) and employing a sparse effective Hamiltonian $\tilde{\mathbf{H}}_k$. The Proxy Hamiltonian $\tilde{\mathbf{H}}_k$ retains exact off-diagonal elements involving $\mathcal{V}_k$ but approximates the $\mathcal{P}_k$-$\mathcal{P}_k$ block diagonally:
\begin{equation}
    (\tilde{\mathbf{H}}_k)_{\mathbf{x}\mathbf{y}} =
    \begin{cases}
        H_{\mathbf{x}\mathbf{y}} & \text{if } \mathbf{x} \in \mathcal{V}_k \text{ or } \mathbf{y} \in \mathcal{V}_k, \\
        H_{\mathbf{x}\mathbf{x}} \delta_{\mathbf{x}\mathbf{y}} & \text{if } \mathbf{x}, \mathbf{y} \in \mathcal{P}_k.
    \end{cases}
    \label{eq:proxy_hamiltonian}
\end{equation}
This approximation in Eq.~\ref{eq:proxy_hamiltonian} avoids the $O(|\mathcal{P}|^2)$ cost of evaluating the full Hamiltonian on the perturbative set. The objective function, defined in Eq.~\ref{eq:proxy_energy}, is the exact Rayleigh quotient of $\tilde{\mathbf{H}}_k$:
\begin{equation}
    E_{\mathrm{proxy}}(\theta) = \frac{\bra{\Psi_{\mathcal{T}}} \tilde{\mathbf{H}}_k \ket{\Psi_{\mathcal{T}}}}{\braket{\Psi_{\mathcal{T}}|\Psi_{\mathcal{T}}}}.
    \label{eq:proxy_energy}
\end{equation}
Since $\tilde{\mathbf{H}}_k$ is Hermitian and $\Omega$ matches the operator's domain, the gradient is well-defined:
\begin{equation}
    \nabla_\theta E_{\mathrm{proxy}} =
    \frac{2}{\braket{\Psi_{\mathcal{T}}|\Psi_{\mathcal{T}}}}
    \sum_{\mathbf{x} \in \mathcal{T}_k}
    \left[
      \sum_{\mathbf{x}' \in \mathcal{T}_k} (\tilde{\mathbf{H}}_k)_{\mathbf{x}\mathbf{x}'} \psi_\theta(\mathbf{x}')
      - E_{\mathrm{proxy}} \psi_\theta(\mathbf{x})
    \right]
    \nabla_\theta \psi_\theta(\mathbf{x}).
    \label{eq:proxy_gradient}
\end{equation}
The gradient in Eq.~\ref{eq:proxy_gradient} provides explicit feedback from amplitudes in $\mathcal{P}_k$, stabilizing optimization in systems with strong coupling to the perturbative set.

\subsubsection{Variational Mode}

The Variational mode restricts the optimization to the variational subspace $\mathcal{V}_k$ ($\Omega = \mathcal{V}_k$), with the effective Hamiltonian defined as the projection $\hat{H}_{\text{eff}} = \hat{P}_{\mathcal{V}} \hat{H} \hat{P}_{\mathcal{V}}$. The objective is the standard variational energy within $\mathcal{V}_k$:
\begin{equation}
    E_{\mathrm{var}}(\theta) = \frac{\bra{\Psi_{\mathcal{V}}} \hat{H} \ket{\Psi_{\mathcal{V}}}}{\braket{\Psi_{\mathcal{V}}|\Psi_{\mathcal{V}}}}.
    \label{eq:variational_energy}
\end{equation}
The gradient corresponds to the projection of the residual vector onto the ansatz tangent space within $\mathcal{V}_k$:
\begin{equation}
    \nabla_\theta E_{\mathrm{var}} =
    \frac{2}{\braket{\Psi_{\mathcal{V}}|\Psi_{\mathcal{V}}}}
    \sum_{\mathbf{x} \in \mathcal{V}_k}
    \left[
      \sum_{\mathbf{x}' \in \mathcal{V}_k} H_{\mathbf{x}\mathbf{x}'} \psi_\theta(\mathbf{x}')
      - E_{\mathrm{var}} \psi_\theta(\mathbf{x})
    \right]
    \nabla_\theta \psi_\theta(\mathbf{x}).
    \label{eq:variational_gradient}
\end{equation}

This mode learns a non-linear compression of the ground state wavefunction within $\mathcal{V}_k$. Since contributions from $\mathcal{P}_k$ are neglected during optimization, the variational objective in Eq.~\ref{eq:variational_energy} and its gradient in Eq.~\ref{eq:variational_gradient} are designed to be coupled with post-optimization perturbative corrections (e.g., PT2) to estimate residual perturbative contributions.

The projected variational objective in Eqs.~\ref{eq:variational_energy} and \ref{eq:variational_gradient} is closely related to the constrained-optimization strategy of Li et al., where an exact truncated NQS energy and its gradient are evaluated within a selected configuration sample\cite{li2024improved}. In the present work, we use this objective as one component of a unified deterministic framework, enabling a direct comparison with the Asymmetric and Proxy target-space formulations in terms of gradient consistency, computational cost, subspace evolution, and residual error sources.

\subsection{Self-Consistent Optimization and Subspace Evolution}
\label{subsec:evolution}

We adopt a self-consistent optimization procedure that interleaves parameter updates with adaptive refinement of the configuration set. This approach decomposes the problem into two alternating steps: minimizing the energy with respect to the ansatz parameters and discretely updating the wavefunction support. The algorithm creates a feedback loop between the effective Hamiltonian, projected onto the current variational subspace, and the neural wavefunction, whose amplitudes determine the set for the subsequent iteration.

\subsubsection{Inner Loop: Parameter Optimization}

In the inner loop, the target set $\mathcal{T}_k$ and the corresponding effective Hamiltonian (e.g., $\tilde{\mathbf{H}}_k$ or $\hat{P}_{\mathcal{V}}\hat{H}\hat{P}_{\mathcal{V}}$) are fixed. The objective is to minimize the deterministic energy functional $E(\theta; \mathcal{T}_k)$ with respect to $\theta$.

Unlike stochastic VMC, the energy and gradients are computed via exact summation over $\mathcal{T}_k$, resulting in a deterministic optimization landscape without sampling noise. Since the effective Hamiltonian and the support $\mathcal{T}_k$ evolve across outer iterations, the network optimizes against a dynamic dataset. This requires the ansatz to learn generalized correlation rules to estimate amplitudes for new configurations entering the set, effectively regularizing the network against overfitting to a static set of configurations.

\subsubsection{Outer Loop: Subspace Evolution}

The outer loop updates the variational basis $\mathcal{V}_k \to \mathcal{V}_{k+1}$ to capture the relevant sector of the Hilbert space. This step functions as a selection operator $\mathcal{S}: \mathcal{T}_k \to \mathcal{V}_{k+1}$, constructing the new variational set based on the amplitude distribution $|\psi_\theta(\mathbf{x})|$ of the current wavefunction. This evolution relies on the neural network's ability to model wavefunction sparsity, assigning large amplitudes to significant configurations even if they currently reside in the perturbative set $\mathcal{P}_k$.

We consider three protocols for subspace selection:
\begin{enumerate}
    \item \textbf{Fixed-Size (Top-$K$) Selection}: The variational set retains the $N_v$ configurations with the largest amplitudes. This imposes an explicit upper bound on memory usage and computational cost, ensuring consistent performance on hardware resources~\cite{liu2025efficient}.
    \item \textbf{Cumulative Mass Selection}: The set expands to include the minimal set of configurations accounting for a target fraction $\gamma \in (0, 1]$ of the total probability mass normalized over $\mathcal{T}_k$.
    \item \textbf{Absolute Threshold Selection}: All configurations with a normalized probability $p(\mathbf{x}) > \epsilon$ are retained. This criterion ensures a uniform resolution of the wavefunction but does not bound the subspace dimension~\cite{li2023nonstochastic, li2024improved}.
\end{enumerate}

In this work, we primarily use the Top-$K$ protocol. This choice decouples the memory footprint from the Hilbert space dimension, providing control over the computational budget and GPU memory utilization.

\subsubsection{Post-Optimization Analysis}

Upon convergence, post-hoc evaluations refine the energy estimate. For the Proxy and Asymmetric modes, the bias from approximate Hamiltonians is removed by evaluating the exact Rayleigh quotient over the final target set $\mathcal{T} = \mathcal{V} \cup \mathcal{P}$. Since the ansatz $\Psi_\theta$ is defined globally, the variational energy can also be estimated via standard VMC sampling~\cite{liu2025efficient}.

In the Variational mode, optimization is confined to $\mathcal{V}$. We estimate the remaining perturbative contribution from the perturbative set using Epstein--Nesbet PT2. A key distinction from linear sCI methods is that the optimized NQS is parameterized by a non-linear network and optimized via gradient descent. Consequently, the final wavefunction may not fully saturate the exact diagonalization limit of the projected Hamiltonian $\hat{P}_{\mathcal{V}}\hat{H}\hat{P}_{\mathcal{V}}$. 

To quantify this internal optimization residual, we define a diagnostic post-hoc diagonalization energy $E_{\mathrm{diag}} = \lambda_{\min}(\mathbf{H}_{\mathcal{VV}})$, evaluated on the final selected configurations. The difference $\Delta_{\mathrm{opt}} = E_{\mathrm{var}} - E_{\mathrm{diag}}$ measures the deviation of the optimized NQS coefficients from the optimal linear CI coefficients within the fixed support $\mathcal{V}$. Detailed diagnostic results are provided in the Supporting Information.

The subsequent perturbative correction estimates contributions associated with this internal residual and with the external coupling to $\mathcal{P}$~\cite{sharma2018stochastic, li2024improved}. The total corrected energy is $E_{\mathrm{tot}} = E_{\mathrm{var}} + \Delta E_{\mathrm{PT2}}$, with:
\begin{equation}
    \Delta E_{\mathrm{PT2}} =
    \sum_{\mathbf{x} \in \mathcal{V}}
    \frac{|\bra{\mathbf{x}}(\hat{H} - E_{\mathrm{var}})\ket{\Psi_{\mathcal{V}}}|^2}{E_{\mathrm{var}} - H_{\mathbf{xx}}}
    +
    \sum_{\mathbf{x} \in \mathcal{P}}
    \frac{|\bra{\mathbf{x}} \hat{H} \ket{\Psi_{\mathcal{V}}}|^2}{E_{\mathrm{var}} - H_{\mathbf{xx}}}.
    \label{eq:pt2_correction}
\end{equation}
The first term in Eq.~\ref{eq:pt2_correction} provides a second-order estimate associated with the non-vanishing residual of the optimized NQS within $\mathcal{V}$, while the second term estimates the perturbative contribution from the external set $\mathcal{P}$.

We also emphasize that minimizing $E_{\mathrm{var}}$ within a selected support does not necessarily minimize the final corrected energy $E_{\mathrm{var}}+\Delta E_{\mathrm{PT2}}$. This mismatch is also observed in selected-CI/PT2 protocols and appears in the present calculations. Therefore, the current Variational+PT2 scheme should be regarded as a two-stage approximation rather than a direct minimization of the corrected total energy. A possible NQS-specific route to reduce this mismatch is to optimize a differentiable surrogate of the corrected energy, or to incorporate PT2-aware scores into the subspace evolution. Such total-energy-driven optimization strategies are left for future work.

\section{Computational Implementation and Acceleration}
\label{sec:implementation}

Implementing a deterministic NQS framework requires managing two distinct workloads: the combinatorial generation of sparse Hamiltonian graphs and the continuous optimization of neural networks via automatic differentiation (AD). To address these differing requirements, we designed a hybrid architecture that separates discrete combinatorial logic (suited for CPUs) from differentiable numerical computation (suited for GPUs).

\subsection{Hybrid Architecture and Memory Management}
\label{subsec:hybrid_arch}

To optimize computational efficiency and memory usage, the framework is structured into three functional layers:

\begin{enumerate}
    \item \textbf{Discrete Combinatorial Kernel}: Evaluates Slater--Condon rules on integer representations of configurations to construct the Hamiltonian connectivity graphs ($\mathbf{H}_{\mathcal{VV}}$, $\mathbf{H}_{\mathcal{VP}}$). This branching-heavy logic is executed on the CPU, exporting sparse operators in Compressed Sparse Row (CSR) format to serve as static constants for the optimization loop.

    \item \textbf{Dynamic Subspace Management}: Acts as a topological filter between the discrete and continuous domains. It applies screening criteria (Section \ref{subsec:acceleration}) to determine the active subspace and transfers only essential topological data (sparse indices and non-zero values) to the optimization engine.

    \item \textbf{Differentiable Optimization Core}: Performs the numerical minimization on the GPU. Using Just-In-Time (JIT) compilation, wavefunction inference and gradient backpropagation are fused into a single computational graph to maximize memory bandwidth utilization.
\end{enumerate}

Deterministic evaluation over large target sets $\mathcal{T}_k$ introduces memory bottlenecks for both sparse matrices and AD activation tensors. We decouple memory storage from high-throughput computation to address this. The sparse Hamiltonian $\mathbf{H}_{\mathcal{TT}}$, which is memory-bound, is stored and processed exclusively in host RAM. The matrix-vector product $\mathbf{v} = \mathbf{H}\mathbf{\psi}$ is evaluated on the CPU, and the data transfer latency is algorithmically masked by concurrent GPU tensor operations. Furthermore, to compute gradients without exceeding GPU memory, the target set $\mathcal{T}_k$ is partitioned into fixed-size micro-batches. Gradients are accumulated sequentially, bounding the peak memory requirement to $O(N_{\text{batch}})$ rather than $O(|\mathcal{T}_k|)$.

\subsection{Acceleration via Heat-Bath Screening}
\label{subsec:acceleration}

The size of the connected subset $\mathcal{C}_k$ scales as $O(N_e^2 N_o^2)$ relative to the variational set $\mathcal{V}_k$. Evaluating the neural ansatz over an unpruned target set $\mathcal{T}_k = \mathcal{V}_k \cup \mathcal{P}_k$ becomes the primary computational bottleneck as the active space grows. We mitigate this by applying a deterministic screening strategy adapted from the Heat-Bath Configuration Interaction (HCI) algorithm\cite{holmes2016efficient, holmes2016heat} to prune the perturbative set $\mathcal{P}_k$.

A candidate configuration $\ket{\mathbf{x}'}$ is included in $\mathcal{T}_k$ only if its perturbative coupling to the variational set satisfies:
\begin{equation}
\max_{\ket{\mathbf{x}} \in \mathcal{V}_k} \left| H_{\mathbf{x}'\mathbf{x}} \psi_\theta(\mathbf{x}) \right| \geq \epsilon_{\mathrm{HB}},
\label{eq:hb_screen}
\end{equation}
where $\psi_\theta(\mathbf{x})$ is the amplitude of the reference state and $\epsilon_{\mathrm{HB}}$ is a numerical threshold.

Evaluating Eq.~\ref{eq:hb_screen} naively incurs an $O(N^4)$ cost per configuration. To accelerate this, we precompute a lookup table during initialization: for each pair of occupied spin-orbitals $(i, j)$, the corresponding pairs of virtual orbitals $(a, b)$ are sorted in descending order of the integral magnitude $|\braket{ij||ab}|$.

During subspace expansion for a configuration $\ket{\mathbf{x}}$, we define a dynamic cutoff $\tau = \epsilon_{\mathrm{HB}} / |\psi_\theta(\mathbf{x})|$. The algorithm iterates through the pre-sorted virtual pairs, enabling an ``early exit'' once $|\braket{ij||ab}| < \tau$. Configurations with negligible coupling are thus never generated. This strategy shifts the complexity of the expansion step from scaling with the total number of orbitals to scaling with the number of significant Hamiltonian entries, ensuring the overall framework remains tractable for large Hilbert spaces.

\subsection{Overview and Optimization Details}
\label{subsec:setup}

The deterministic optimization workflow is summarized in Algorithm~\ref{alg:det_optimization}. Unless otherwise noted, molecular Hamiltonians are constructed using one- and two-body integrals obtained from Restricted Hartree--Fock (RHF) calculations via the PySCF package~\cite{sun2018pyscf, sun2020recent, sun2015libcint}. The initial variational set $\mathcal{V}_0$ consists solely of the Hartree--Fock reference configuration.

The backflow correction $\Delta\mathbf{\Phi}_\theta$ is parameterized by a Multi-Layer Perceptron (MLP) with two hidden layers of 256 units. No pre-training is used. The reference matrix $\mathbf{\Phi}^0$ is initialized from the Hartree--Fock occupation matrix, with a small Gaussian perturbation of standard deviation $10^{-3}$ added for symmetry breaking. In the JAX/Flax implementation, the embedding and hidden layers use standard random initializers with zero biases. The final output projection uses a small variance-scaling initialization with zero bias, keeping $\Delta\mathbf{\Phi}_\theta(\mathbf{x}) \approx \mathbf{0}$ at initialization. Thus, the starting wavefunction is close to the Hartree--Fock reference, while all parameters of the backflow correction remain trainable. The numerical optimization is implemented using the JAX library~\cite{jax2018github} with double-precision (fp64) arithmetic.

To realize the hybrid architecture described in Section~\ref{subsec:hybrid_arch}, the discrete combinatorial kernel is implemented in C++ and interfaced with Python via \texttt{nanobind}~\cite{nanobind} to ensure efficient integer operations. The differentiable inner loop is Just-In-Time (JIT) compiled into a unified execution graph. Within this compiled loop, sparse matrix-vector products are routed to the host CPU utilizing optimized SciPy~\cite{virtanen2020scipy} routines, while memory-efficient gradient accumulation over large target sets is managed sequentially via JAX control flow primitives.

Optimization employs the AdamW algorithm with weight decay. The learning rate follows a cosine decay schedule, decreasing from $10^{-3}$ to $5 \times 10^{-5}$ over 800 steps. Calculations involve $N_{\text{outer}} = 30$ outer iterations for subspace evolution. In each outer iteration, we perform $N_{\text{inner}} = 1000$ gradient descent steps. Large target set evaluations use a micro-batch size of 8192 to manage GPU memory. The heat-bath screening threshold for the perturbative set is set to $\epsilon_{\text{HB}} = 10^{-6}$. All benchmarks were performed on a single NVIDIA A100 (80GB) GPU with 8 Intel Xeon Gold 8358 CPU cores and 128GB RAM.

\begin{algorithm}[htbp]
\caption{Deterministic Optimization of Neural Network Quantum States (Variational Mode)}
\label{alg:det_optimization}
\renewcommand{\baselinestretch}{1.2}\selectfont
\begin{algorithmic}[1]
\State \textbf{Input:} Hamiltonian integrals (FCIDUMP), Ansatz $\Psi_\theta$, Hyperparameters ($N_{\text{outer}}$, $N_{\text{inner}}$, $\epsilon_{\text{HB}}$, $K$)
\State \textbf{Output:} Optimized parameters $\theta^*$, final energy $E_{\text{total}}$
\Statex
\State Initialize parameters $\theta$ and variational set $\mathcal{V}_0 \leftarrow \{ \ket{\Phi_{\text{HF}}} \}$
\For{$k = 1$ to $N_{\text{outer}}$} \Comment{\textbf{Outer Loop: Subspace Evolution}}
    \Statex
    \State \textbf{1. Subspace Expansion (Layer 0)}
    \State \quad Transfer amplitudes $\psi_\theta(\mathcal{V}_{k-1})$ to host
    \State \quad Generate screened perturbative set $\mathcal{P}_k$ via heat-bath criteria:
    \State \qquad $\mathcal{P}_k \leftarrow \{ \ket{\mathbf{x}'} \notin \mathcal{V}_{k-1} \mid \max_{\mathbf{x} \in \mathcal{V}_{k-1}} |H_{\mathbf{x}'\mathbf{x}} \psi_\theta(\mathbf{x})| \ge \epsilon_{\text{HB}} \}$
    \State \quad Construct sparse Hamiltonian block $\mathbf{H}_{\mathcal{VV}}$
    \Statex
    \State \textbf{2. Differentiable Parameter Optimization (Layer 2)}
    \State \quad Transfer configurations $\mathcal{V}_{k-1}$ to device memory
    \State \quad \textbf{Compile} optimization step (fused forward/backward pass):
    \For{$t = 1$ to $N_{\text{inner}}$} \Comment{\textbf{Inner Loop}}
        \State \qquad Compute amplitudes $\psi_\theta(\mathcal{V}_{k-1})$ via serialized inference
        \State \qquad Compute matrix-vector product $\mathbf{v} = \mathbf{H}_{\mathcal{VV}} \mathbf{\psi}$ (Host-Resident)
        \State \qquad $\mathcal{L}(\theta) \leftarrow E_{\text{var}} = \frac{\bra{\Psi_{\mathcal{V}}} \hat{H} \ket{\Psi_{\mathcal{V}}}}{\braket{\Psi_{\mathcal{V}}|\Psi_{\mathcal{V}}}}$
        \State \qquad Compute gradients $\nabla_\theta \mathcal{L}$ via AD
        \State \qquad Update $\theta \leftarrow \text{Optimizer}(\theta, \nabla_\theta \mathcal{L})$
    \EndFor
    \Statex
    \State \textbf{3. Subspace Pruning (Layer 1)}
    \State \quad Evaluate wavefunction amplitudes on full target set $\mathcal{T}_k$
    \State \quad $\mathcal{V}_k \leftarrow \text{Selector}(\mathcal{T}_k, K, \text{key}=|\psi_\theta|)$
\EndFor
\Statex
\State \textbf{4. Post-Optimization Correction}
\State Compute $E_{\text{var}}$ on final set $\mathcal{V}_{N_{\text{outer}}}$
\State Compute perturbative correction $\Delta E_{\text{PT2}}$
\State \Return $\theta^*, E_{\text{total}} = E_{\text{var}} + \Delta E_{\text{PT2}}$
\end{algorithmic}
\end{algorithm}

\section{Results and Discussion}
\label{sec:results}

We present numerical results to validate the accuracy and efficiency of the deterministic framework. The analysis begins with an evaluation of the optimization modes and convergence properties on the water molecule. We then benchmark ground-state energies for small molecules against FCI and stochastic NQS methods, followed by an investigation of bond dissociation curves for \ce{H2O} and \ce{N2}. Finally, we apply the framework to the strongly correlated chromium dimer (\ce{Cr2}) and analyze the wall-time scaling of the hybrid implementation.

\subsection{Evaluation of Optimization Modes and Dynamics}
\label{subsec:mode_comparison}

\begin{figure}[h]
  \centering
  \includegraphics[width=1.0\textwidth]{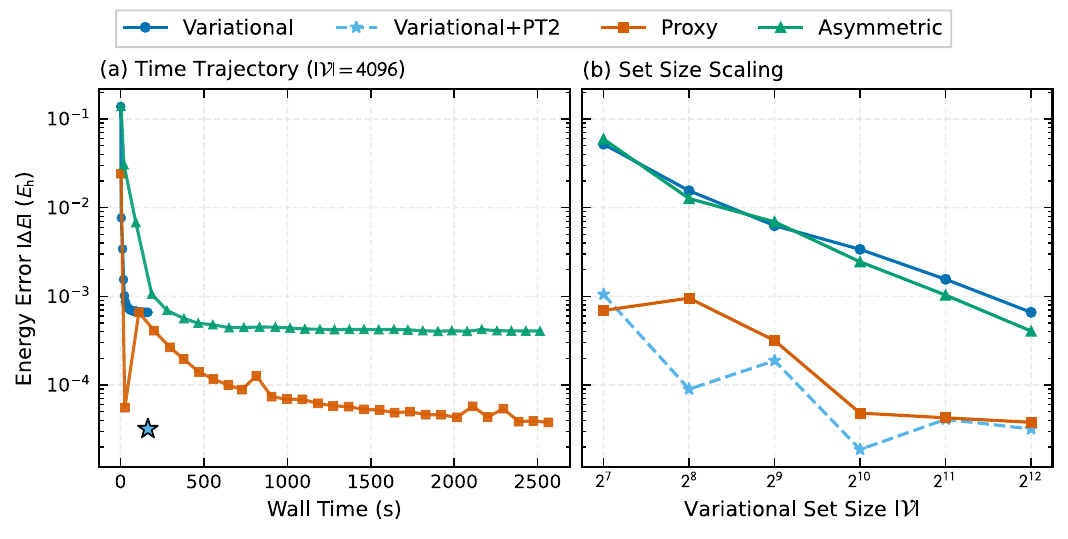}
  \caption{Convergence of deterministic objectives for \ce{H2O} (6-31G). (a) Energy error $|\Delta E|$ relative to FCI versus wall time for a fixed variational set $|\mathcal{V}|=4096$. (b) Final energy error versus variational set size $|\mathcal{V}|$.}
  \label{fig:h2o_convergence}
\end{figure}

We first evaluate the convergence characteristics of the Variational, Asymmetric, and Proxy modes using \ce{H2O} in the 6-31G basis set. Figure~\ref{fig:h2o_convergence} displays the energy error relative to the FCI limit as a function of wall time and variational set size $|\mathcal{V}|$.

The wall-time trajectories in Figure~\ref{fig:h2o_convergence}(a) distinguish the computational costs of the three approaches. The Variational mode achieves the lowest cost per inner-loop iteration because wavefunction evaluation and gradient backpropagation are strictly confined to the variational set $\mathcal{V}$, avoiding repeated target-set evaluations over the connected configurations during gradient optimization. While the pure Variational energy is bounded by the representational capacity of $\mathcal{V}$ (solid blue line), the addition of the second-order Epstein--Nesbet perturbation correction (Variational+PT2, dashed blue line) accounts for the majority of the remaining perturbative-space contribution in this test without increasing the inner-loop optimization cost.

Regarding the target-space formulations ($\mathcal{T} = \mathcal{V} \cup \mathcal{P}$), the Proxy mode (orange line) yields significantly lower errors than the Asymmetric mode (green line). The stagnation of the Asymmetric mode near the uncorrected Variational result indicates that evaluating Hamiltonian connections into $\mathcal{P}$ is ineffective without a consistent gradient signal. The gradient estimator in the Asymmetric mode neglects the parameter dependence of amplitudes in the perturbative set ($\nabla_\theta \psi(\mathbf{x})$ for $\mathbf{x} \in \mathcal{P}$), resulting in a non-conservative vector field. In contrast, the Proxy mode minimizes the exact Rayleigh quotient of a Hermitian effective Hamiltonian $\tilde{\mathbf{H}}$ defined on $\mathcal{T}$. This ensures mathematical consistency between the energy functional and its gradient, allowing the optimization to effectively incorporate contributions from the perturbative set.

The error scaling with respect to subspace size $|\mathcal{V}|$ (Figure~\ref{fig:h2o_convergence}(b)) demonstrates that the Proxy mode and the Variational+PT2 strategy yield comparable accuracy. This agreement indicates that, for this test case, the post-hoc PT2 correction accounts for a similar portion of the perturbative-space energy contribution as the Proxy formulation, although the corrected energy remains non-variational. However, the Proxy mode requires evaluating gradients over the full target set $\mathcal{T}$ at every step, whereas the PT2 correction is a non-iterative calculation performed only upon convergence. Given its theoretical consistency and superior computational efficiency, we adopt the Variational+PT2 protocol as the default protocol for the subsequent benchmarks.

Having established the Variational+PT2 objective, we briefly address the algorithmic choices for the self-consistent procedure (detailed in the Supporting Information). Analysis of the inner-loop convergence reveals a system dependence: weakly correlated systems (\ce{H2O}) plateau rapidly after subspace updates, whereas strongly correlated systems (\ce{Cr2}) benefit from sustained optimization to navigate complex energy landscapes (Figures S1 and S2). To provide a conservative and uniform protocol across diverse benchmarks without \textit{a priori} tuning, we adopted a fixed schedule of 30 outer loops and 1000 inner steps. Diagnostic ablation studies (Figure S3 and Table S1) demonstrate that the final total energy of the Variational+PT2 scheme is largely insensitive to this inner/outer allocation in the tested regime, supporting the use of a relatively infrequent outer-loop update schedule to amortize the CPU overhead of combinatorial graph generation. Furthermore, while various subspace selection criteria yield comparable accuracy when compared at similar subspace sizes (Figure S4), the Top-$K$ strategy is exclusively employed as it strictly bounds the memory footprint, ensuring predictable GPU performance for large-scale systems.

\subsection{Benchmarks on Molecular Systems}
\label{subsec:molecular_benchmark}

We evaluate the scalability and accuracy of the deterministic framework by computing the ground-state energies of \ce{Li2O}, \ce{C2H4O}, and \ce{C3H8}. These systems encompass Hilbert spaces ranging from $10^7$ to $10^{12}$ configurations. Table~\ref{tab:benchmark_energies} summarizes the calculated energies alongside results from FCI, CCSD(T), and existing NQS benchmarks.

A key difference between this work and previous deterministic or semi-deterministic NQS approaches (e.g., SC-RBM \cite{li2023nonstochastic, li2024improved}, NNBF \cite{liu2024nnbf, liu2025efficient}) lies in the optimization domain. Some previous deterministic or semi-deterministic NQS approaches incorporate the connected set $\mathcal{C}$ or a target set $\mathcal{T}$ into the gradient estimator to capture dynamic correlation during training. Depending on the precise objective-gradient pairing, such target-space formulations may introduce additional computational cost or gradient-consistency issues. In contrast, the present Variational Mode restricts both wavefunction evaluation and gradient backpropagation strictly to the variational set $\mathcal{V}$. This makes the dominant inner-loop cost scale with the selected variational support rather than the larger target set, and defines a Hermitian variational objective for the projected Hamiltonian $\hat{P}_\mathcal{V} \hat{H} \hat{P}_\mathcal{V}$. This approach effectively decouples the problem: the neural ansatz optimizes the static correlation within a compact $\mathcal{V}$, while the residual perturbative contribution is estimated via the post-hoc PT2 correction.

For \ce{Li2O}, strictly variational minimization within a compact set proves efficient. Using $|\mathcal{V}| = 2\,048$, the variational energy ($-87.891\,996$ Ha) is lower than the SC-RBM result obtained with a similar approach ($-87.891\,8$ Ha) \cite{li2023nonstochastic}. This suggests that a closed, mathematically consistent variational objective provides an efficient optimization route, although direct comparisons also depend on the ansatz, optimizer, and implementation details. As $|\mathcal{V}|$ expands to $131\,072$, the total energy (Variational+PT2) converges to the FCI limit within micro-Hartree accuracy.

For \ce{C2H4O}, where the Hilbert space exceeds $10^9$, we compare our results against the autoregressive MADE\cite{zhao2023scalable} and NAQS\cite{barrett2022autoregressive} architecture. Our variational energy using 2\,048 configurations ($-151.104\,900$ Ha) is lower than the MADE benchmark ($-151.092\,983$ Ha). While autoregressive models rely on unbiased sampling, the deterministic Top-$K$ selection imposes a strong inductive bias that prioritizes configurations based on amplitude magnitude. This allows the ansatz to identify dominant configurations in highly sparse wavefunctions without the variance associated with stochastic sampling.

The \ce{C3H8} molecule represents a regime ($10^{12}$ configurations) where exact enumeration is intractable. Comparison with recent distributed FCI benchmarks \cite{gao2024distributed} shows systematic convergence. Expanding $|\mathcal{V}|$ from $131\,072$ to $524\,288$ reduces the total energy error relative to FCI from $0.96$ mHa to $0.47$ mHa. The results demonstrate that while the variational optimization captures the multi-configurational nature of the reference state, the perturbative correction is essential for accounting for the remaining perturbative contribution in large basis sets.

\begin{table}[htbp]
\centering
\caption{Ground-state energies (in Hartree) calculated by our deterministic framework compared with previous NQS benchmarks and FCI/CCSD(T) references. $|\mathcal{V}|$ denotes the size of the variational set. For our method, `Variational Energy' refers to the exact expectation value $E = \bra{\Psi_\mathcal{V}} \hat{H} \ket{\Psi_\mathcal{V}} / \braket{\Psi_\mathcal{V}|\Psi_\mathcal{V}}$, and `Total Energy' includes the post-optimization PT2 correction. Methods are sorted by variational energy (descending) within each system.}
\label{tab:benchmark_energies}
\begin{tabular}{l l r r}
\toprule
Method & {$|\mathcal{V}|$} & {Variational Energy} & {Total Energy} \\
\midrule
\multicolumn{4}{l}{\textit{\ce{Li2O} (STO-3G, 14e, 15o, $|H| \approx 4.14 \times 10^7$)}} \\
\midrule
FCI & {-} & -87.892693 & {-} \\
CCSD(T) & {-} & {-} & -87.893089 \\
\addlinespace[0.5ex]
MADE \cite{zhao2023scalable} & {-} & -87.885637 & {-} \\
\textbf{This work} & 512 &  -87.887889 &  -87.892584 \\
SC-RBM \cite{li2023nonstochastic} & {-} & -87.8918 & {-} \\
\textbf{This work} & 2 048 &  -87.891996 &  -87.892680 \\
NAQS \cite{barrett2022autoregressive} & {-} & -87.892433 & {-} \\
\textbf{This work} & 8 192 &  -87.892541 &  -87.892689 \\
Transformer \cite{shang2025solving} & {-} & -87.8926 & {-} \\
\textbf{This work} & 32 768 &  -87.892646 &  -87.892691 \\
NNBF \cite{liu2024nnbf} & {-} & -87.892662 & {-} \\
\textbf{This work} & 131 072 &  -87.892662 &  -87.892692 \\
\midrule
\multicolumn{4}{l}{\textit{\ce{C2H4O} (STO-3G, 24e, 19o, $|H| \approx 2.54 \times 10^9$)}} \\
\midrule
FCI & {-} & -151.123570 & {-} \\
CCSD(T) & {-} & {-} & -151.122748 \\
\addlinespace[0.5ex]
\textbf{This work} & 512 &  -151.047686 &  -151.115556 \\
MADE \cite{zhao2023scalable} & {-} & -151.092983 & {-} \\
\textbf{This work} & 2 048 &  -151.104900 &  -151.121939 \\
\textbf{This work} & 8 192 &  -151.114672 &  -151.122729 \\
\textbf{This work} & 32 768 &  -151.119006 &  -151.123226 \\
NAQS \cite{barrett2022autoregressive} & {-} & -151.120486 & {-} \\
\textbf{This work} & 131 072 &  -151.121237 &  -151.123361 \\
\textbf{This work} & 524 288 &  -151.122393 &  -151.123470 \\
Transformer \cite{shang2025solving} & {-} & -151.1228 & {-} \\
NNBF \cite{liu2024nnbf} & {-} & -151.123357 & {-} \\
\midrule
\multicolumn{4}{l}{\textit{\ce{C3H8} (STO-3G, 26e, 23o, $|H| \approx 1.31 \times 10^{12}$)}} \\
\midrule
FCI \cite{gao2024distributed} & {-} & -117.100123 & {-} \\
CCSD(T) & {-} & {-} & -117.099709 \\
\addlinespace[0.5ex]
\textbf{This work} & 512 &  -116.967731 &  -117.088963 \\
\textbf{This work} & 2 048 &  -117.068360 &  -117.095751 \\
\textbf{This work} & 8 192 &  -117.082970 &  -117.097458 \\
\textbf{This work} & 32 768 &  -117.087785 &  -117.098339 \\
\textbf{This work} & 131 072 &  -117.092824 &  -117.099158 \\
\textbf{This work} & 524 288 &  -117.096772 &  -117.099649 \\
\bottomrule
\end{tabular}
\end{table}

\subsection{Dissociation of \texorpdfstring{\ce{H2O}}{H2O} and \texorpdfstring{\ce{N2}}{N2}}
\label{subsec:dissociation}

We computed potential energy surfaces (PES) for the symmetric dissociation of \ce{H2O} and the bond breaking of \ce{N2} using the cc-pVDZ basis set. Reference ground-state energies were obtained from FCI data by Olsen et al.\cite{olsen1996full} for \ce{H2O}, and DMRG ($M=4000$)\cite{chan2004state} and CDFCI\cite{wang2019coordinate} benchmarks for \ce{N2}.

\begin{figure}[h]
  \centering
  \includegraphics[width=1.0\textwidth]{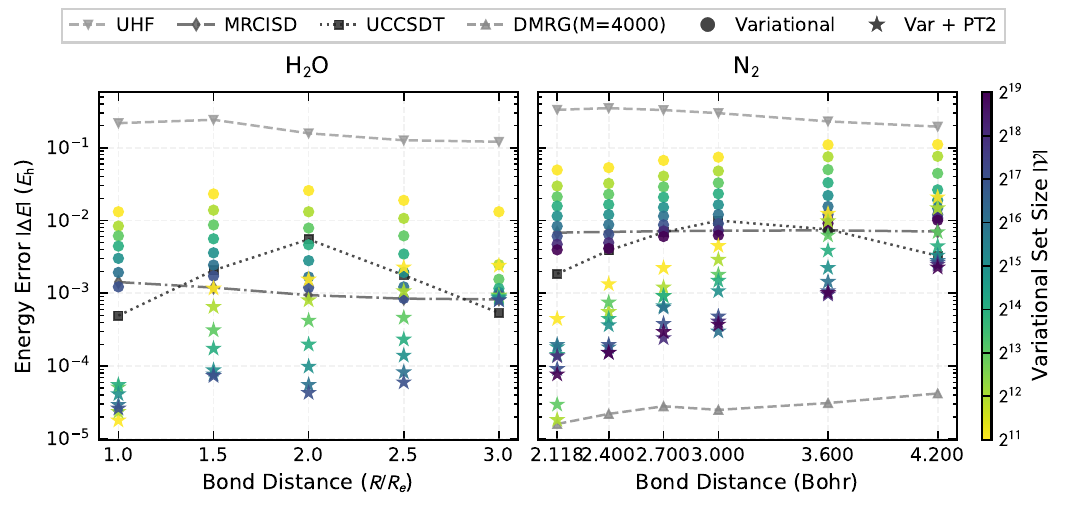}
  \caption{Absolute energy errors $|\Delta E|$ relative to FCI/CDFCI benchmarks for the symmetric dissociation of \ce{H2O} (left) and \ce{N2} (right) in the cc-pVDZ basis set. Results are shown for the Variational mode (circles) and the Variational+PT2 scheme (stars). The color gradient indicates the size of the variational set $|\mathcal{V}|$, ranging from $2^{11}$ (yellow) to $2^{17}$ (dark purple). Baselines including UHF, MRCISD, UCCSDT, and DMRG are plotted for comparison.}
  \label{fig:dissociation}
\end{figure}

Figure~\ref{fig:dissociation} presents the absolute energy errors relative to the benchmarks. For \ce{H2O} (left panel), the single-reference UCCSDT method diverges from the FCI limit as the O--H bonds stretch, with errors increasing by approximately two orders of magnitude at $3.0 R/R_e$. This reflects the inadequacy of the single-configuration reference in the bond-breaking regime. The present neural backflow ansatz, optimized via the Variational mode and augmented with second-order perturbation theory (Variational+PT2), maintains consistent accuracy across the dissociation coordinate. With a variational set of $|\mathcal{V}| = 2^{17}$, the Variational+PT2 energies remain within $10^{-4}$~Ha of the FCI reference, yielding lower errors than both UCCSDT and MRCISD at large bond distances.

The dissociation of \ce{N2} (right panel) involves breaking a triple bond, introducing strong static correlation. The UCCSDT baseline exhibits a non-monotonic error distribution with a maximum deviation around 3.0 Bohr. This behavior arises from the symmetry-broken UHF reference, where spin contamination and the multireference character of the wavefunction compromise accuracy in the intermediate coupling regime.

As shown in Figure~\ref{fig:dissociation} (right), increasing $|\mathcal{V}|$ from $2^{13}$ to $2^{17}$ systematically reduces the variational energy error. The inclusion of the PT2 correction accounts for a large portion of the remaining perturbative contribution, particularly near the equilibrium geometry where it reduces errors to below the MRCISD baseline. However, at the dissociation limit (e.g., 4.2 Bohr), the total energy error increases to approximately 2.2 mHa. Post-hoc diagonalization diagnostics (see Supporting Information) reveal that the internal optimization residual $\Delta_{\mathrm{opt}}$ grows noticeably in this strongly correlated regime, reaching $\sim 0.62$ mHa for $|\mathcal{V}|=131\,072$. This diagnostic gap accounts for only part of the total deviation. The remaining deviation is likely associated with a combination of finite selected support, limitations of amplitude-based selection for resolving many near-degenerate configurations, and the reduced reliability of a post-hoc second-order correction when the external space contains important statically correlated configurations.

\subsection{Chromium Dimer}

\begin{figure}[htbp]
    \centering
    \includegraphics[width=\linewidth]{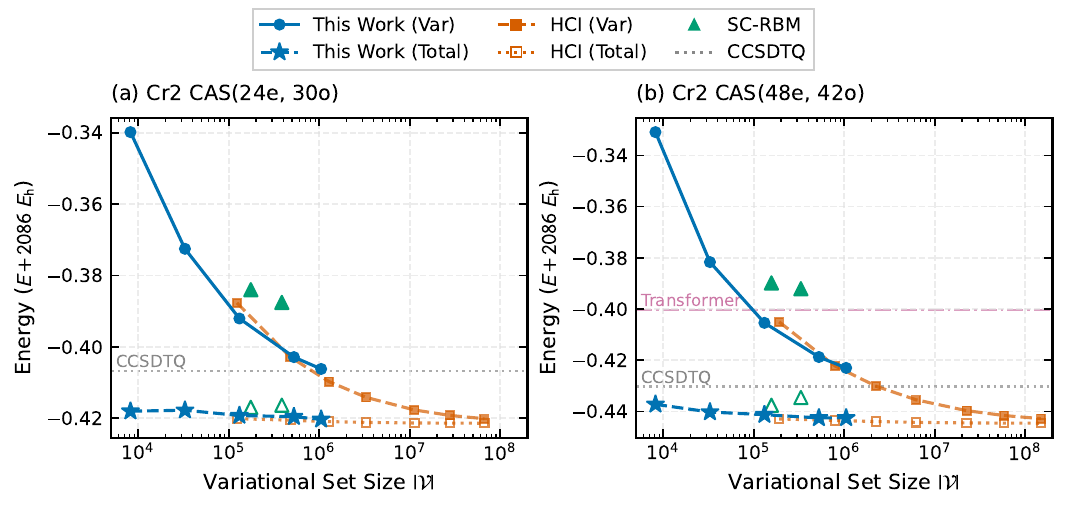}
    \caption{Convergence of variational (circles) and PT2-corrected total energies (stars) for \ce{Cr2} as a function of the variational set size $|\mathcal{V}|$. (a) CAS(24e, 30o) and (b) CAS(48e, 42o). Reference data include HCI (orange squares), SC-RBM (green triangles), Transformer (pink dashed line), and CCSDTQ (gray dotted line).}
    \label{fig:cr2_convergence}
\end{figure}

The chromium dimer (\ce{Cr2}) serves as a benchmark for balancing strong static and dynamic correlation. We computed the ground-state energy at a bond length of $1.5$ \AA{} using the Ahlrichs SV basis set. To ensure comparability with Selected CI (HCI) \cite{li2020accurate}, DMRG \cite{olivares2015ab}, and stochastic NQS methods (SC-RBM \cite{li2024improved}, Transformer \cite{kan2025nnqs}), all calculations utilize identical molecular orbitals derived from CASSCF(12e, 12o). Results for the CAS(24e, 30o) and CAS(48e, 42o) active spaces are summarized in Table~\ref{tab:cr2_energies_combined} and Figure~\ref{fig:cr2_convergence}.

In the CAS(24e, 30o) system (System I, Hilbert space $\sim 7.5 \times 10^{15}$), the deterministic framework achieves lower variational energies with smaller subspace sizes compared to stochastic baselines. As shown in Figure~\ref{fig:cr2_convergence}(a), at $|\mathcal{V}| = 8,192$, the PT2-corrected total energy ($-0.418002$ Ha) is lower than the CCSDTQ reference ($-0.406696$ Ha). This indicates that single-reference CCSDTQ is not a sufficiently stringent multireference benchmark for \ce{Cr2}; a more appropriate assessment relies on deviations from HCI and DMRG. Comparing the variational energy at similar subspace sizes, our result at $|\mathcal{V}| \approx 1.31 \times 10^5$ ($-0.392029$ Ha) is approximately $8$ mHa lower than the SC-RBM result ($-0.38396$ Ha), which employs a support of $\sim 1.75 \times 10^5$. Upon expanding $|\mathcal{V}|$ to $10^6$, the total energy converges to $-0.420278$ Ha, consistent with DMRG and FCIQMC limits.

For the larger CAS(48e, 42o) active space (System II, Hilbert space $> 10^{23}$), single-reference CCSDTQ overestimates the energy by $\sim 14$ mHa relative to DMRG. Figure~\ref{fig:cr2_convergence}(b) shows that the deterministic ansatz yields variational energies lower than reported Transformer results. The final total energy of $-0.442491$ Ha is within about $2$ mHa of the best available many-reference benchmarks from HCI and DMRG ($\approx -0.4445$ Ha).

The variational energy trajectory generally follows the HCI curve, suggesting that amplitude-based selection captures a physically relevant subset of configurations. However, the slope of the NQS curve becomes shallower than that of HCI as the subspace size increases. This divergence reflects the methodological difference between the two approaches: HCI performs an exact diagonalization within the selected subspace, whereas NQS relies on non-linear parameter optimization.

Post-hoc subspace diagonalization diagnostics (detailed in the Supporting Information) confirm this behavior. At the largest subspace size ($|\mathcal{V}| \approx 10^6$), the internal optimization residual $\Delta_{\mathrm{opt}}$ reaches approximately 1.7 mHa and 1.9 mHa for System I and System II, respectively. This residual is of the same order as the remaining energy deviation from the selected-CI references, suggesting that coefficient optimization within the selected support is a non-negligible error source in massive Hilbert spaces. While the internal PT2 term in Eq.~\ref{eq:pt2_correction} perturbatively estimates a correction for this residual, the final accuracy remains bounded by the combined effects of the non-linear optimization limits and the perturbative screening threshold.

\begin{table}[htbp]
\centering
\caption{Ground-state energies of the \ce{Cr2} dimer in different active spaces. All energies are reported as $E + 2086$ Ha.}
\label{tab:cr2_energies_combined}
\begin{tabular}{l l r r}
\toprule
Method & {$|\mathcal{V}|$} & {Variational Energy} & {Total Energy} \\
\midrule
\multicolumn{4}{l}{\textit{System I: CAS(24e, 30o), $|H| \approx 7.48 \times 10^{15}$}} \\
\midrule
CCSDTQ \cite{olivares2015ab} & {-} & {-} & -0.406696 \\
\textbf{This work} & 8 192 &  -0.339793 &  -0.418002 \\
\textbf{This work} & 32 768 &  -0.372492 &  -0.417737 \\
SC-RBM \cite{li2024improved} & 174 963 & -0.38396 & -0.416920 \\
SC-RBM \cite{li2024improved} & 385 554 & -0.38752 & -0.416310 \\
HCI \cite{li2020accurate} & 123 144 & -0.387661 & -0.420094 \\
\textbf{This work} & 131 072 &  -0.392029 &  -0.419212 \\
\textbf{This work} & 524 288 &  -0.402847 &  -0.419550 \\
HCI \cite{li2020accurate} & 480 138 & -0.402856 & -0.420541 \\
\textbf{This work} & 1 048 576 &  -0.406173 &  -0.420278 \\
HCI \cite{li2020accurate} & 66 679 956 & -0.420114 & -0.421375 \\
DMRG \cite{sharma2012spin} & {-} & -0.420525 & -0.421156 \\
FCIQMC \cite{booth2014linear} & {-} & {-} & -0.421200 \\
\midrule
\multicolumn{4}{l}{\textit{System II: CAS(48e, 42o), $|H| \approx 1.25 \times 10^{23}$}} \\
\midrule
CCSDTQ \cite{olivares2015ab} & {-} & {-} & -0.430244 \\
\textbf{This work} & 8 192 &  -0.330898 &  -0.437190 \\
\textbf{This work} & 32 768 &  -0.381646 &  -0.440195 \\
SC-RBM \cite{li2024improved} & 156 594 & -0.38978 & -0.437530 \\
SC-RBM \cite{li2024improved} & 331 397 & -0.39203 & -0.434520 \\
SC-Transformer \cite{kan2025nnqs} & {-} & -0.40043 & {-} \\
HCI \cite{li2020accurate} & 190 937 & -0.405000 & -0.442899 \\
\textbf{This work} & 131 072 &  -0.405377 &  -0.441221 \\
\textbf{This work} & 524 288 &  -0.418717 &  -0.442491 \\
HCI \cite{li2020accurate} & 787 919 & -0.422163 & -0.443463 \\
\textbf{This work} & 1 048 576 &  -0.422971 & -0.442491  \\
HCI \cite{li2020accurate} & 2 237 828 & -0.430093 & -0.443908 \\
HCI \cite{li2020accurate} & 148 589 206 & -0.442773 & -0.444560 \\
DMRG \cite{olivares2015ab} & {-} & -0.443334 & -0.444478 \\
\bottomrule
\end{tabular}
\end{table}

\subsection{Computational Efficiency and Complexity Analysis}
\label{subsec:complexity}

\begin{figure}[htbp]
  \centering
  \includegraphics[width=1.0\textwidth]{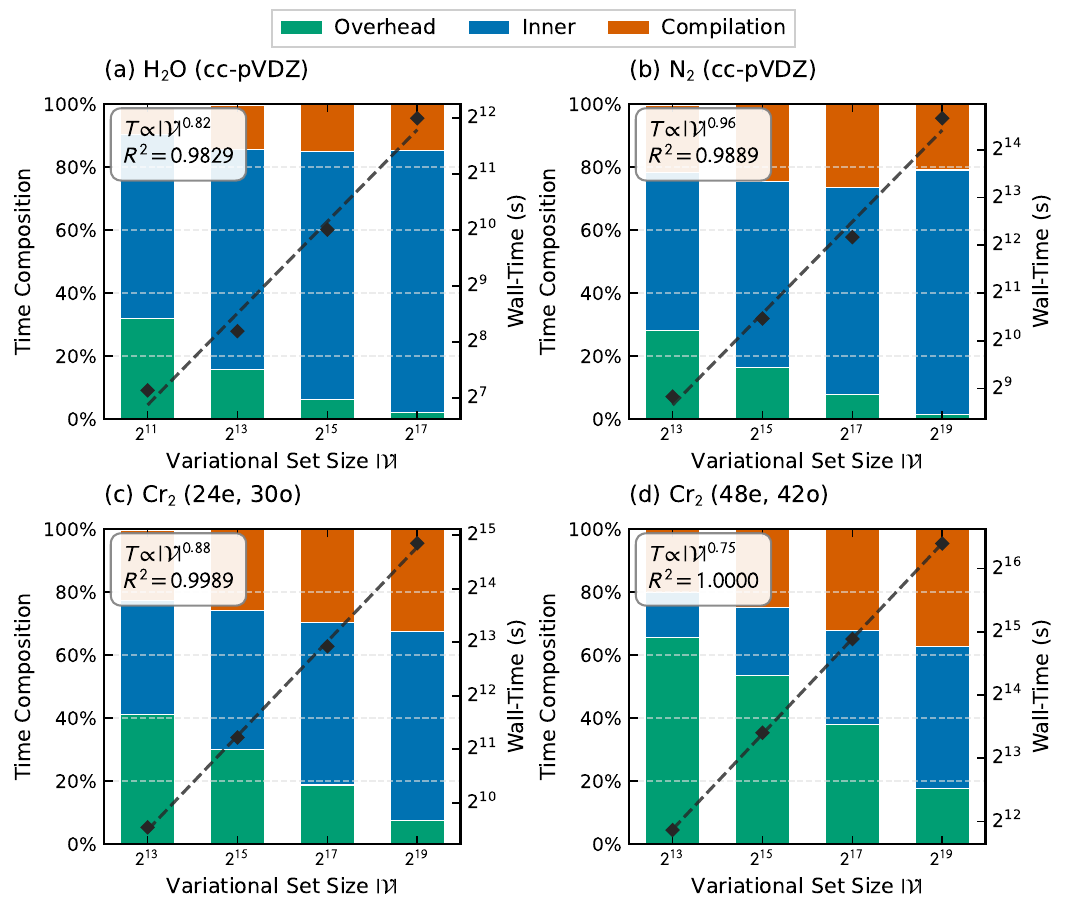}
  \caption{Wall-time decomposition of a single outer-loop iteration versus the variational set size $|\mathcal{V}|$ for (a) \ce{H2O}, (b) \ce{N2}, (c) \ce{Cr2} (24e, 30o), and (d) \ce{Cr2} (48e, 42o). The stacked bars (left axis) represent the relative percentage of each computational phase, while the diamond markers (right axis) track the total wall-time in seconds. The total scaling follows a power law $T \propto |\mathcal{V}|^\alpha$, with fitted exponents $\alpha$ provided in the insets.}
  \label{fig:timing_benchmark}
\end{figure}

We analyze the scaling performance of the deterministic framework by decomposing the total wall-time into three computational phases:
(1) \textit{Compilation} (Orange): Construction of the sparse Hamiltonian graph and connected set $\mathcal{C}$ via the C++ kernel;
(2) \textit{Inner Loop} (Blue): JAX-compiled optimization steps, performing wavefunction evaluation and gradient accumulation on the GPU;
(3) \textit{Overhead} (Green): Subspace management tasks, including candidate scoring, sorting, and host-device data transfer.
Figure~\ref{fig:timing_benchmark} details the time decomposition for \ce{H2O}, \ce{N2}, and \ce{Cr2} active spaces as a function of the variational set size $|\mathcal{V}|$.

The total wall-time follows a power law scaling $T \propto |\mathcal{V}|^\alpha$ across all benchmarked systems, with exponents ranging from $\alpha = 0.75$ to $\alpha = 0.88$. This empirical sub-linear scaling indicates that the marginal computational cost per configuration decreases as the subspace expands, reflecting the efficiency of batched operations on large datasets. However, the distribution of computational effort depends heavily on the electronic structure complexity.

For the \ce{H2O} and \ce{N2} systems (Figure~\ref{fig:timing_benchmark}(a,b)), the workflow is compute-bound. The \textit{Inner Loop} dominates the total runtime (50--80\%), as the sparse Hamiltonian graph remains relatively compact. In this regime, performance is limited primarily by the arithmetic intensity of the neural network inference and the GPU memory bandwidth during gradient accumulation.

In contrast, the strongly correlated \ce{Cr2} system (Figure~\ref{fig:timing_benchmark}(c,d)) exhibits a different profile due to the dense connectivity of the active space. For the large CAS(48e, 42o), the \textit{Overhead} and \textit{Compilation} phases constitute the majority of the wall-time at small subspace sizes ($>60\%$ at $|\mathcal{V}|=2^{13}$). This behavior occurs because the size of the connected set $|\mathcal{C}|$ scales rapidly with the number of open shells, making the CPU-based integer logic for graph generation and screening the primary bottleneck.

As $|\mathcal{V}|$ increases, however, the relative contribution of the \textit{Inner Loop} rises (e.g., to $\sim40\%$ in Figure~\ref{fig:timing_benchmark}(d)). This trend suggests that the algorithmic overhead for subspace management effectively saturates, shifting the workload back towards the highly parallelizable GPU operations in the large-scale limit. Additionally, the \textit{Compilation} time remains controlled despite the massive Hilbert space, confirming that the heat-bath screening efficiently prunes the Hamiltonian connectivity graph. The observed scaling exponents ($\alpha < 0.9$) demonstrate that the hybrid architecture successfully balances discrete combinatorial logic on the host with differentiable linear algebra on the device, maintaining tractability for variational sets exceeding $10^5$ configurations.

\section{Conclusion and Outlook}
\label{sec:conclusion}

We have presented a deterministic optimization framework for NQS, reformulating the electronic Schrödinger equation as a non-linear optimization problem over dynamically adaptive configuration subspaces. By replacing stochastic gradient estimation with exact summation, this approach eliminates the sampling noise and slow mixing associated with MCMC in discrete Fock spaces. Through a comparison of the Asymmetric, Proxy, and Variational modes, we analyzed the gradient consistency and computational costs of different deterministic objectives. Our results indicate that restricting the neural ansatz optimization to a compact variational set ($\mathcal{V}$), combined with a post-hoc PT2 correction, provides an efficient and internally consistent approach for optimizing the selected variational component while estimating residual correlation through a post-hoc PT2 correction.

Benchmarks on small molecules and bond dissociations show that the method optimizes important static-correlation components within the selected support and estimates residual perturbative contributions, alleviating some limitations of single-reference coupled-cluster descriptions in bond-breaking regimes. For the strongly correlated \ce{Cr2} in large active spaces [CAS(48e, 42o)], the framework maintains accuracy comparable to the sCI references while exhibiting empirical sub-linear wall-time scaling with respect to the set size over the tested range ($T \propto |\mathcal{V}|^\alpha, \alpha < 1$). These findings indicate that neural backflow ansätze can efficiently approximate complex wavefunctions in Hilbert spaces exceeding $10^{23}$, provided the dominant amplitudes are localized within a treatable subspace. Furthermore, post-hoc diagonalization diagnostics show that in large, strongly correlated subspaces, the gradient-optimized NQS may not fully reach the exact subspace diagonalization limit, leaving millihartree-scale internal relaxation errors.

However, the current time-to-solution remains higher than mature sCI and DMRG solvers. The primary bottleneck lies in the high arithmetic intensity of neural network tensor contractions compared to the optimized sparse matrix-vector products used in linear CI methods. Additionally, the efficiency of deterministic subspace expansion relies on the sparsity of the ground state in the chosen orbital basis. For systems with dense entanglement or significant delocalization, the rapid growth of the required variational support $|\mathcal{V}|$ may diminish the computational advantages of this formulation.

Future work will address these limitations through the following extensions:

\begin{itemize}
    \item \textbf{Orbital Optimization:} As sparsity is basis-dependent, integrating orbital optimization (e.g., via simultaneous gradient updates) is necessary to minimize the variational set size $|\mathcal{V}|$ required for target accuracy, thereby compacting the wavefunction representation\cite{smith2017cheap, yao2021orbital}.

    \item \textbf{Symmetry Adaptation:} The current ansatz implies but does not strictly enforce total spin ($S^2$) or point-group symmetries. Explicitly incorporating these constraints via symmetry-projected networks or penalty functions will reduce spin contamination, particularly in open-shell transition metals\cite{jimenez2012projected, luo2023gauge}.

    \item \textbf{Advanced Ansatz Architectures:} While the MLP backflow provides a generic baseline, future work can explore architectures with stronger physical priors or higher expressivity. This includes incorporating explicit physical constraints via Pfaffian and Jastrow structures, or hybrid Neuro-Tensor Network states to improve data efficiency\cite{chen2025neural, wu2025hybrid, du2025neuralized, li2025representational}. Furthermore, integrating geometry-aware architectures, such as Graph Neural Networks, or leveraging the long-range dependency modeling of Transformers, could enhance the capture of non-local correlations in larger basis sets\cite{shang2025solving, liang2021hybrid}.

    \item \textbf{Optimization Algorithms and Strategies:} The availability of noise-free gradients within the deterministic framework facilitates the application of advanced optimization techniques to address ill-conditioned curvature. This includes curvature-aware algorithms such as exact SR and its low-rank variants (e.g., minSR)\cite{sorella1998green, sorella2001generalized, chen2024empowering}. Additionally, the absence of sampling variance allows for precise gradient-based convergence control and dynamic optimization strategies, such as adaptive learning rate scheduling and automated hyperparameter stepping.

    \item \textbf{Hybrid Deterministic--Stochastic Schemes:} To estimate perturbative contributions from the vast external space without excessive memory costs, a hybrid approach may be employed\cite{sharma2017semistochastic, liu2025efficient}. This would combine deterministic optimization of the strongly correlated core with stochastic sampling of the perturbative tail, balancing variational rigor with the scalability of Monte Carlo integration.
\end{itemize}

In summary, this framework formulates NQS optimization within a rigorous linear algebra context, treating the network as a differentiable function approximator rather than a probability sampler. By decoupling ansatz design from sampling challenges, it offers a distinct pathway for applying deep learning to high-precision ab initio quantum chemistry.

\section*{Supporting Information}
Supporting Information. Exact gradient derivation for the asymmetric objective; inner-loop convergence traces for \ce{H2O} and \ce{Cr2}; inner/outer-loop allocation benchmarks; subspace selector comparison; post-hoc subspace diagonalization diagnostics for \ce{N2} and \ce{Cr2}; detailed numerical data and wall-time decomposition.

\begin{acknowledgement}
The author thanks the Supercomputing Center of the University of Science and Technology of China for providing the computational resources used in this work. The author is also grateful to Qiang Fu for helpful discussions and support. The NetKet codebase is acknowledged as a learning and inspiration resource for the development of the software framework used in this study~\cite{netket3:2022,netket2:2019}. Additionally, the author acknowledges the use of generative AI for language polishing and text refinement to improve the clarity and readability of the manuscript. No specific funding sources are declared for this manuscript.
\end{acknowledgement}

\section*{Data Availability}

The data that support the findings of this study, including molecular geometries and detailed energy results, are available within the article and its Supporting Information. Any additional raw data are available from the corresponding author upon reasonable request.

\section*{Code Availability}

The source code for the deterministic NQS framework, including the hybrid CPU-GPU implementation and example scripts, is publicly available on GitHub at \href{https://github.com/wsmxcz/DetNQS}{https://github.com/wsmxcz/DetNQS}. This repository also contains the molecular Hamiltonian integrals (in FCIDUMP format) used for the benchmarks presented in this study.

\bibliography{reference}

\clearpage

% ---------- SI numbering ----------
\setcounter{page}{1}
\setcounter{section}{0}
\setcounter{figure}{0}
\setcounter{table}{0}
\setcounter{equation}{0}

\renewcommand{\thepage}{S\arabic{page}}
\renewcommand{\thesection}{S\arabic{section}}
\renewcommand{\thefigure}{S\arabic{figure}}
\renewcommand{\thetable}{S\arabic{table}}
\renewcommand{\theequation}{S\arabic{equation}}

\begin{center}
{\Large\bfseries Supporting Information\par}
\vspace{0.5em}
{\large A Deterministic Framework for Neural Network Quantum States in Quantum Chemistry\par}
\vspace{1.5em}
Zheng Che\\
Hefei National Research Center for Physical Sciences at the Microscale,\\
University of Science and Technology of China, Hefei 230026, China\\
Email: wsmxcz@gmail.com
\end{center}

\vspace{2em}

\section*{Contents}

\noindent
\hyperref[sec:asym_grad]{S1. Exact Gradient of the Asymmetric Energy Functional}
\dotfill \pageref{sec:asym_grad}

\noindent
\hyperref[sec:si-optimization-dynamics]{S2. Optimization Dynamics and Algorithmic Choices}
\dotfill \pageref{sec:si-optimization-dynamics}

\noindent
\hyperref[sec:si-subspace-diag]{S3. Post-Hoc Subspace Diagonalization Diagnostics}
\dotfill \pageref{sec:si-subspace-diag}

\noindent
\hyperref[sec:si-data]{S4. Detailed Numerical Data}
\dotfill \pageref{sec:si-data}

\clearpage

\section{Exact Gradient of the Asymmetric Energy Functional}
\label{sec:asym_grad}

The Asymmetric energy functional corresponds to a deterministic evaluation of a VMC-like local-energy average, normalized over the variational set $\mathcal{V}$ while retaining Hamiltonian couplings from $\mathcal{V}$ into the target set $\mathcal{T}=\mathcal{V}\cup\mathcal{P}$:
\begin{equation}
    E_{\mathrm{asym}}(\theta)
    =
    \frac{N(\theta)}{D(\theta)}
    =
    \frac{\bra{\Psi_{\mathcal{V}}} \hat{H} \ket{\Psi_{\mathcal{T}}}}
         {\braket{\Psi_{\mathcal{V}}|\Psi_{\mathcal{V}}}}
    =
    \frac{\sum_{\mathbf{x}\in\mathcal{V}} \psi_\theta(\mathbf{x})^2 E_{\mathrm{loc}}(\mathbf{x})}
         {\sum_{\mathbf{x}\in\mathcal{V}} \psi_\theta(\mathbf{x})^2},
    \label{eq:si-asym-energy}
\end{equation}
where
\begin{equation}
    E_{\mathrm{loc}}(\mathbf{x})
    =
    \sum_{\mathbf{y}\in\mathcal{T}}
    H_{\mathbf{x}\mathbf{y}}
    \frac{\psi_\theta(\mathbf{y})}{\psi_\theta(\mathbf{x})}.
    \label{eq:si-asym-local-energy}
\end{equation}

Let $\boldsymbol{\psi}_{\mathcal{V}}$ and $\boldsymbol{\psi}_{\mathcal{P}}$ denote the vectors of amplitudes on $\mathcal{V}$ and $\mathcal{P}$, respectively. The numerator and denominator can be written as
\begin{align}
    N(\theta)
    &=
    \boldsymbol{\psi}_{\mathcal{V}}^{T}
    \mathbf{H}_{\mathcal{VV}}
    \boldsymbol{\psi}_{\mathcal{V}}
    +
    \boldsymbol{\psi}_{\mathcal{V}}^{T}
    \mathbf{H}_{\mathcal{VP}}
    \boldsymbol{\psi}_{\mathcal{P}},
    \label{eq:si-asym-numerator}
    \\
    D(\theta)
    &=
    \boldsymbol{\psi}_{\mathcal{V}}^{T}
    \boldsymbol{\psi}_{\mathcal{V}}.
    \label{eq:si-asym-denominator}
\end{align}
Here $\mathbf{H}_{\mathcal{VP}}$ denotes the Hamiltonian block with rows in $\mathcal{V}$ and columns in $\mathcal{P}$, and $\mathbf{H}_{\mathcal{PV}}=\mathbf{H}_{\mathcal{VP}}^{T}$ for a real symmetric Hamiltonian. Products involving $\nabla_\theta \boldsymbol{\psi}_{\mathcal{V}}$ and $\nabla_\theta \boldsymbol{\psi}_{\mathcal{P}}$ should be understood as contractions with the corresponding Jacobians with respect to the network parameters.

Differentiating the quotient gives
\begin{equation}
    \nabla_\theta E_{\mathrm{asym}}
    =
    \frac{\nabla_\theta N(\theta)
    -
    E_{\mathrm{asym}}\nabla_\theta D(\theta)}
    {D(\theta)}.
    \label{eq:si-asym-quotient}
\end{equation}
Using the symmetry of $\mathbf{H}_{\mathcal{VV}}$, the numerator derivative is
\begin{equation}
    \nabla_\theta N
    =
    \left(
    2\mathbf{H}_{\mathcal{VV}}\boldsymbol{\psi}_{\mathcal{V}}
    +
    \mathbf{H}_{\mathcal{VP}}\boldsymbol{\psi}_{\mathcal{P}}
    \right)^T
    \nabla_\theta \boldsymbol{\psi}_{\mathcal{V}}
    +
    \left(
    \mathbf{H}_{\mathcal{PV}}\boldsymbol{\psi}_{\mathcal{V}}
    \right)^T
    \nabla_\theta \boldsymbol{\psi}_{\mathcal{P}},
    \label{eq:si-asym-numerator-gradient}
\end{equation}
whereas
\begin{equation}
    \nabla_\theta D
    =
    2\boldsymbol{\psi}_{\mathcal{V}}^{T}
    \nabla_\theta \boldsymbol{\psi}_{\mathcal{V}}.
    \label{eq:si-asym-denominator-gradient}
\end{equation}
Substitution into Eq.~\ref{eq:si-asym-quotient} yields the exact gradient
\begin{equation}
    \nabla_\theta E_{\mathrm{asym}}
    =
    \frac{1}{D}
    \left[
    \left(
    2\mathbf{H}_{\mathcal{VV}}\boldsymbol{\psi}_{\mathcal{V}}
    +
    \mathbf{H}_{\mathcal{VP}}\boldsymbol{\psi}_{\mathcal{P}}
    -
    2E_{\mathrm{asym}}\boldsymbol{\psi}_{\mathcal{V}}
    \right)^T
    \nabla_\theta \boldsymbol{\psi}_{\mathcal{V}}
    +
    \left(
    \mathbf{H}_{\mathcal{PV}}\boldsymbol{\psi}_{\mathcal{V}}
    \right)^T
    \nabla_\theta \boldsymbol{\psi}_{\mathcal{P}}
    \right].
    \label{eq:si-asym-exact-gradient}
\end{equation}

The commonly used VMC-like gradient estimator for this asymmetric objective is
\begin{equation}
    \mathbf{g}_{\mathrm{asym}}
    =
    2\sum_{\mathbf{x}\in\mathcal{V}}
    \rho_{\mathcal{V}}(\mathbf{x})
    \left[
    E_{\mathrm{loc}}(\mathbf{x})-E_{\mathrm{asym}}
    \right]
    \nabla_\theta \ln |\psi_\theta(\mathbf{x})|,
    \label{eq:si-asym-truncated-gradient}
\end{equation}
where
\begin{equation}
    \rho_{\mathcal{V}}(\mathbf{x})
    =
    \frac{\psi_\theta(\mathbf{x})^2}
         {\sum_{\mathbf{z}\in\mathcal{V}}\psi_\theta(\mathbf{z})^2}.
    \label{eq:si-asym-rho}
\end{equation}
In matrix form, Eq.~\ref{eq:si-asym-truncated-gradient} is
\begin{equation}
    \mathbf{g}_{\mathrm{asym}}
    =
    \frac{2}{D}
    \left(
    \mathbf{H}_{\mathcal{VV}}\boldsymbol{\psi}_{\mathcal{V}}
    +
    \mathbf{H}_{\mathcal{VP}}\boldsymbol{\psi}_{\mathcal{P}}
    -
    E_{\mathrm{asym}}\boldsymbol{\psi}_{\mathcal{V}}
    \right)^T
    \nabla_\theta \boldsymbol{\psi}_{\mathcal{V}}.
    \label{eq:si-asym-truncated-matrix}
\end{equation}

Comparing Eq.~\ref{eq:si-asym-exact-gradient} with Eq.~\ref{eq:si-asym-truncated-matrix}, the difference between the exact gradient and the truncated VMC-like estimator is
\begin{equation}
    \nabla_\theta E_{\mathrm{asym}}
    -
    \mathbf{g}_{\mathrm{asym}}
    =
    \frac{1}{D}
    \left[
    \left(
    \mathbf{H}_{\mathcal{PV}}\boldsymbol{\psi}_{\mathcal{V}}
    \right)^T
    \nabla_\theta \boldsymbol{\psi}_{\mathcal{P}}
    -
    \left(
    \mathbf{H}_{\mathcal{VP}}\boldsymbol{\psi}_{\mathcal{P}}
    \right)^T
    \nabla_\theta \boldsymbol{\psi}_{\mathcal{V}}
    \right].
    \label{eq:si-asym-gradient-difference}
\end{equation}
Therefore, the truncated estimator is not the exact gradient of $E_{\mathrm{asym}}$ unless the coupling to $\mathcal{P}$ vanishes or the omitted target-space derivative terms cancel accidentally. This mismatch is the origin of the gradient inconsistency discussed in the main text. 

\clearpage

\section{Optimization Dynamics and Algorithmic Choices}
\label{sec:si-optimization-dynamics}

The optimization dynamics of Neural Network Quantum States (NQS) within the deterministic framework are influenced by the electronic structure of the target system and the allocation of computational resources during the self-consistent procedure. In this section, we analyze these effects by examining the inner-loop convergence traces, the computational budget allocation, and the subspace selection criteria.

\subsection{System-Dependent Convergence Behavior}

The convergence behavior during the inner-loop parameter optimization varies depending on the correlation strength of the molecule. To illustrate this, we track the variational energy evaluated at each inner-loop gradient step for a weakly correlated system (\ce{H2O} near equilibrium) and a strongly correlated system (\ce{Cr2}).

\begin{figure}[htbp]
\centering
\includegraphics[width=0.85\textwidth]{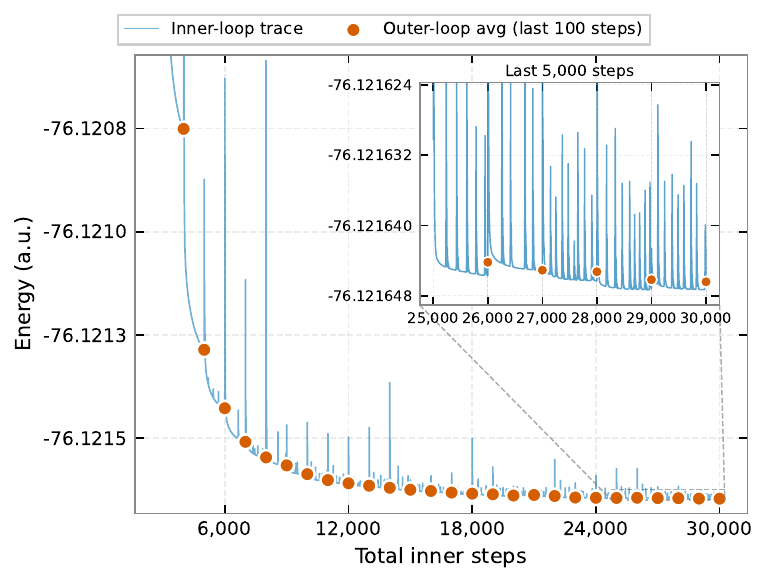}
\caption{
Inner-loop optimization history for \ce{H2O} (6-31G) with a variational set size of $|\mathcal{V}|=4096$. The energy exhibits transient spikes upon outer-loop subspace updates, followed by rapid relaxation and a subsequent plateau.
}
\label{fig:si-h2o-history}
\end{figure}

\begin{figure}[htbp]
\centering
\includegraphics[width=0.85\textwidth]{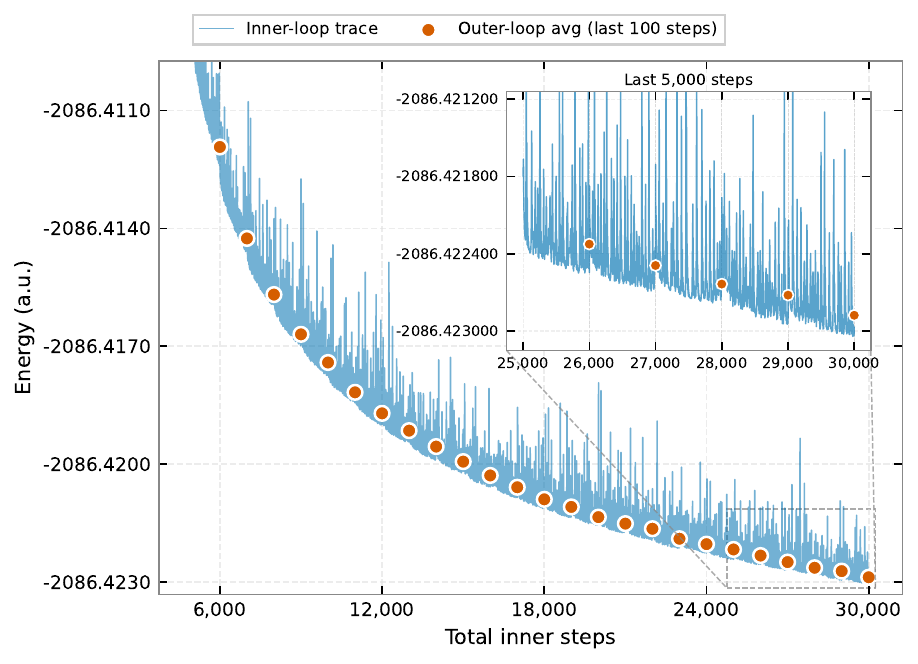}
\caption{
Inner-loop optimization history for the strongly correlated \ce{Cr2} dimer (48e, 42o) with $|\mathcal{V}|=1\,048\,576$. The energy exhibits a continuous descent over thousands of steps without clear saturation.
}
\label{fig:si-cr2-history}
\end{figure}

As shown in Figure~\ref{fig:si-h2o-history}, for the \ce{H2O} molecule, the variational energy exhibits a characteristic pattern where outer-loop updates induce discontinuities or transient energy spikes, followed by rapid relaxation. In this weakly correlated regime, executing a large number of inner-loop steps (e.g., 1000 steps) quickly leads to a plateau, yielding diminishing returns in energy minimization. 

Conversely, Figure~\ref{fig:si-cr2-history} demonstrates the dynamics for the \ce{Cr2} dimer in a massive active space. Due to the strong multireference character and the high dimensionality of the selected subspace ($>10^6$ configurations), the optimization landscape is highly complex. The energy continues to decrease steadily throughout the 1000 inner steps, showing no clear saturation within the chosen inner-loop window. This observation motivates the use of a uniform, conservative inner-loop budget for benchmarking diverse chemical systems: while automated early-stopping mechanisms would accelerate computations for simpler molecules, a sustained inner-loop budget benefits the optimization of network parameters in strongly correlated regimes.

\subsection{Inner- and Outer-Loop Allocation}

\begin{figure}[htbp]
\centering
\includegraphics[width=0.65\textwidth]{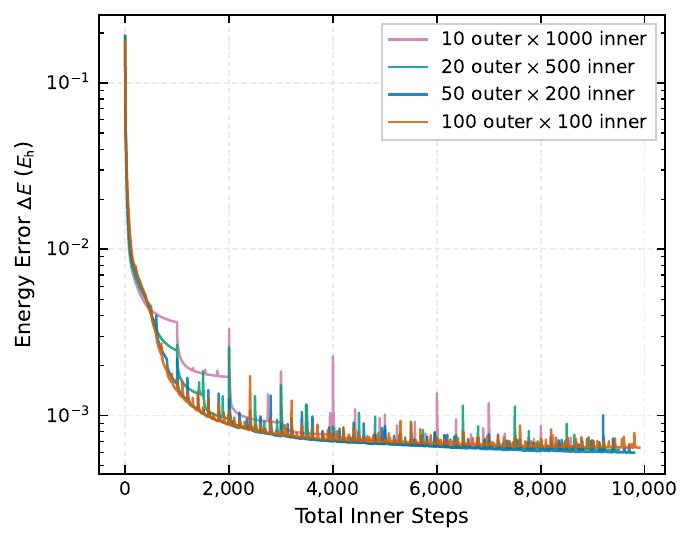}
\caption{
Energy error $\Delta E$ relative to FCI for different inner/outer loop allocations under a fixed budget of 10,000 total gradient steps for \ce{H2O} (6-31G).
}
\label{fig:si-outer-inner}
\end{figure}

Given a fixed total budget of gradient updates, the allocation between outer loops (subspace evolution) and inner loops (parameter optimization) affects both the wall-time efficiency and the final accuracy. We performed an ablation study on \ce{H2O} (6-31G) with a fixed total budget of 10,000 gradient steps and a subspace size of $|\mathcal{V}|=4096$. The results are summarized in Figure~\ref{fig:si-outer-inner} and Table~\ref{tab:si-outer-inner}.

\begin{table}[htbp]
\centering
\caption{
Computational cost and final energies for different outer/inner-loop allocations and optimization modes. The total gradient budget is fixed at 10,000 steps for \ce{H2O} (6-31G) with $|\mathcal{V}|=4096$.
}
\label{tab:si-outer-inner}
\begin{tabular}{l l r r}
\toprule
Mode & Allocation & Wall Time / s & $E_{\mathrm{total}}$ / Ha \\
\midrule
\multirow{4}{*}{Variational + PT2} 
& 100 outer $\times$ 100 inner & 133.9 & -76.121664 \\
& 50 outer $\times$ 200 inner  & 80.5  & -76.121707 \\
& 20 outer $\times$ 500 inner  & 50.1  & -76.121683 \\
& 10 outer $\times$ 1000 inner & 39.5  & -76.121662 \\
\midrule
\multirow{4}{*}{Proxy} 
& 100 outer $\times$ 100 inner & 1044.6 & -76.122174 \\
& 50 outer $\times$ 200 inner  & 915.6  & -76.122224 \\
& 20 outer $\times$ 500 inner  & 810.6  & -76.122211 \\
& 10 outer $\times$ 1000 inner & 717.1  & -76.122238 \\
\midrule
\multirow{4}{*}{Asymmetric} 
& 100 outer $\times$ 100 inner & 1013.1 & -76.121809 \\
& 50 outer $\times$ 200 inner  & 899.9  & -76.121805 \\
& 20 outer $\times$ 500 inner  & 795.7  & -76.121876 \\
& 10 outer $\times$ 1000 inner & 706.0  & -76.121905 \\
\bottomrule
\end{tabular}
\end{table}

The data in Table~\ref{tab:si-outer-inner} show that reducing the frequency of outer loops (e.g., 10 outer $\times$ 1000 inner) significantly decreases the total wall time. This reduction occurs because subspace expansion and Hamiltonian graph generation involve combinatorial logic that acts as a substantial CPU overhead. Despite the different optimization trajectories shown in Figure~\ref{fig:si-outer-inner}, the Variational+PT2 scheme yields final total energies that are largely insensitive to the allocation strategy under this specific setup (varying by only $\sim 4.5 \times 10^{-5}$ Ha). We note that while the 50 outer $\times$ 200 inner allocation achieved the lowest energy, the 10 outer $\times$ 1000 inner allocation was the fastest. Therefore, fixing a relatively large inner-loop budget is a practical choice aimed at balancing wall-time efficiency and benchmark consistency, rather than a universally optimal strategy. Furthermore, under this common update budget, the target-space modes (Proxy and Asymmetric) remain substantially more expensive computationally.

\subsection{Subspace Selection Criteria}
\label{sec:si-selector}

The outer loop updates the variational basis $\mathcal{V}_k$ based on the amplitude distribution of the current wavefunction. We analyze three distinct selection criteria: Top-$K$, Cumulative Mass, and Absolute Threshold. The Cumulative Mass selector retains the minimal set of determinants satisfying $\sum_{i=1}^{N_v} p_i \ge \gamma$, where $p_i = |\psi_\theta(\mathbf{x}_i)|^2 / \sum_{\mathbf{x} \in \mathcal{T}} |\psi_\theta(\mathbf{x})|^2$. The Absolute Threshold selector retains all configurations where the normalized probability $p_i > \epsilon$.

\begin{figure}[htbp]
\centering
\includegraphics[width=0.65\textwidth]{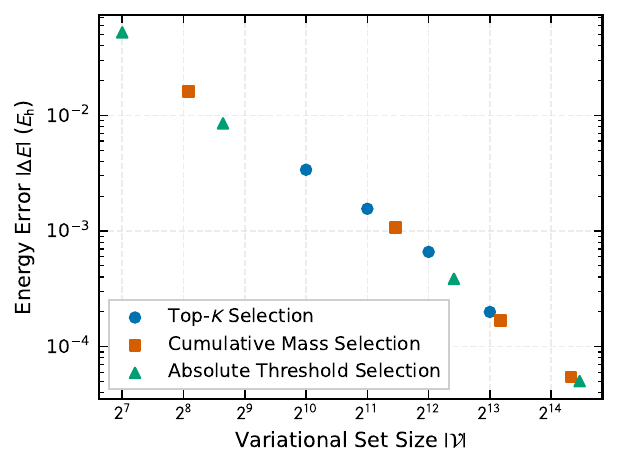}
\caption{
Final variational energy error $|E_{\mathrm{var}}-E_{\mathrm{FCI}}|$ versus the resulting variational set size $|\mathcal{V}|$ for different subspace selection criteria (\ce{H2O} 6-31G).
}
\label{fig:si-selector}
\end{figure}

\begin{table}[htbp]
\centering
\caption{
Performance and final subspace dimensions for different selection criteria on \ce{H2O} (6-31G). Calculations were performed using 30 outer loops and 1000 inner steps.
}
\label{tab:si-selector-scan}
\begin{tabular}{l l r r r}
\toprule
Selector & Parameter & Final $|\mathcal{V}|$ & $E_{\mathrm{var}}$ / Ha & Wall Time / s \\
\midrule
\multirow{4}{*}{Top-$K$} 
& $K=1024$ & 1 024 & -76.118919 & 107.6 \\
& $K=2048$ & 2 048 & -76.120748 & 120.5 \\
& $K=4096$ & 4 096 & -76.121645 & 164.4 \\
& $K=8192$ & 8 192 & -76.122105 & 253.8 \\
\midrule
\multirow{4}{*}{Cumulative Mass} 
& $\gamma=0.999$    & 271    & -76.106198 & 73.3 \\
& $\gamma=0.9999$   & 2 822  & -76.121230 & 152.4 \\
& $\gamma=0.99999$  & 9 230  & -76.122137 & 445.9 \\
& $\gamma=0.999999$ & 20 540 & -76.122250 & 679.4 \\
\midrule
\multirow{4}{*}{Absolute Threshold} 
& $\epsilon=10^{-4}$ & 128    & -76.070147 & 44.3 \\
& $\epsilon=10^{-6}$ & 400    & -76.113778 & 87.8 \\
& $\epsilon=10^{-8}$ & 5 454  & -76.121919 & 225.9 \\
& $\epsilon=10^{-10}$& 22 609 & -76.122255 & 681.5 \\
\bottomrule
\end{tabular}
\end{table}

As illustrated in Figure~\ref{fig:si-selector} and Table~\ref{tab:si-selector-scan}, the variational energy decreases systematically with the retained subspace size $|\mathcal{V}|$. Notably, the energy reduction closely follows this trend regardless of the specific selection rule applied. In this test, the final accuracy is more strongly correlated with the retained subspace size than with the specific selector identity.

The choice among these criteria represents a practical engineering trade-off between physical adaptability and computational predictability. From a physical perspective, the Cumulative Mass and Absolute Threshold criteria are more natural: they dynamically adapt to the intrinsic sparsity of different molecular wavefunctions without requiring \textit{a priori} knowledge of the optimal subspace size. In contrast, the Top-$K$ approach generally requires trial-and-error tuning, as the appropriate $K$ needed to reach a specific accuracy varies across different chemical systems.

However, the dynamic subspace sizes generated by the Mass and Threshold criteria are difficult to predict. For instance, tightening the threshold $\epsilon$ from $10^{-8}$ to $10^{-10}$ leads to a more than fourfold increase in $|\mathcal{V}|$ (from 5,454 to 22,609). In large-scale calculations, such unpredictable growth can cause sudden memory spikes and out-of-memory errors on GPU devices. By strictly bounding the memory footprint and the computational cost per iteration, the Top-$K$ criterion ensures stable and predictable hardware execution. 

We note that the current fixed-budget and Top-$K$ settings serve as a diagnostic baseline for consistent benchmarking. Implementing hybrid dynamic strategies---such as combining an adaptive threshold with a hard memory cap, or using PT2-based scoring for configuration selection---represents a valuable direction for balancing physical adaptability and optimization efficiency in the future.

\clearpage

\section{Post-Hoc Subspace Diagonalization Diagnostics}
\label{sec:si-subspace-diag}

As discussed in Section 2.4.3 of the main text, the non-linear gradient optimization of the NQS ansatz may not fully saturate the exact diagonalization limit of the selected subspace $\mathcal{V}$, particularly in massive Hilbert spaces with strong multireference character. To isolate the internal relaxation error within a fixed selected support, we perform post-hoc exact diagonalizations on the final selected configurations.

Here, we denote the optimized NQS variational energy on the selected support as $E_{\mathrm{opt}}$ (corresponding to $E_{\mathrm{var}}$ in the main text) for the diagnostic comparison. We define the diagnostic diagonalization energy as $E_{\mathrm{diag}} = \lambda_{\min}(\mathbf{H}_{\mathcal{VV}})$. The optimization residual is then given by:
\begin{equation}
    \Delta_{\mathrm{opt}} = E_{\mathrm{opt}} - E_{\mathrm{diag}}.
\end{equation}
This metric quantifies the deviation of the NQS coefficients from the optimal linear CI coefficients within the fixed support $\mathcal{V}$. Note that this procedure serves exclusively as a diagnostic tool for error attribution; it does not replace the self-consistent Variational+PT2 protocol used for production calculations. The diagonalized coefficients are not used to update the neural-network parameters or to define the reported production energies.

Table~\ref{tab:si-diag-n2} presents the diagnostics for the \ce{N2} molecule at both the equilibrium (2.118 Bohr) and stretched (4.200 Bohr) geometries. The data reveal that $\Delta_{\mathrm{opt}}$ generally increases with the subspace dimension $|\mathcal{V}|$, especially beyond $|\mathcal{V}|=32\,768$. Furthermore, at the stretched geometry, where static correlation dominates, the residual is notably larger (e.g., 2.76 mHa at $|\mathcal{V}|=524\,288$) compared to the equilibrium geometry (1.22 mHa). This indicates that, under the present ansatz and optimization protocol, coefficient optimization becomes more difficult in the stretched multireference regime.

\begin{table}[htbp]
\centering
\caption{
Post-hoc subspace diagonalization diagnostics for \ce{N2} (cc-pVDZ) at equilibrium (2.118 Bohr) and stretched (4.200 Bohr) geometries. $\Delta_{\mathrm{opt}}$ is reported in mHa for clarity.
}
\label{tab:si-diag-n2}
\begin{tabular}{l r r r r}
\toprule
Geometry & $|\mathcal{V}|$ & $E_{\mathrm{opt}}$ / Ha & $E_{\mathrm{diag}}$ / Ha & $\Delta_{\mathrm{opt}}$ / mHa \\
\midrule
\multirow{7}{*}{$d = 2.118$ Bohr}
& 8 192   & -109.260943 & -109.260943 & 0.001 \\
& 16 384  & -109.266361 & -109.266362 & 0.001 \\
& 32 768  & -109.270646 & -109.270655 & 0.009 \\
& 65 536  & -109.273813 & -109.273892 & 0.079 \\
& 131 072 & -109.276065 & -109.276345 & 0.280 \\
& 262 144 & -109.277462 & -109.278137 & 0.675 \\
& 524 288 & -109.278217 & -109.279435 & 1.218 \\
\midrule
\multirow{7}{*}{$d = 4.200$ Bohr}
& 8 192   & -108.925608 & -108.925610 & 0.002 \\
& 16 384  & -108.943700 & -108.943702 & 0.002 \\
& 32 768  & -108.951043 & -108.951135 & 0.092 \\
& 65 536  & -108.955324 & -108.955541 & 0.217 \\
& 131 072 & -108.958120 & -108.958741 & 0.621 \\
& 262 144 & -108.959330 & -108.960824 & 1.493 \\
& 524 288 & -108.960053 & -108.962815 & 2.762 \\
\bottomrule
\end{tabular}
\end{table}

Table~\ref{tab:si-diag-cr2} presents the corresponding diagnostics for the \ce{Cr2} dimer in two active spaces. For the massive CAS(48e, 42o) space, the optimization residual reaches 1.93 mHa at $|\mathcal{V}| \approx 10^6$. This magnitude is consistent with the observed gap between the NQS variational energy curve and the HCI variational curve presented in the main text. It implies that while amplitude-based selection captures a physically relevant subset of configurations, saturating the exact diagonalization limit within these massive strongly correlated subspaces remains an ongoing challenge for non-linear NQS optimization.

\begin{table}[htbp]
\centering
\caption{
Post-hoc subspace diagonalization diagnostics for the \ce{Cr2} dimer. Energies are reported as $E + 2086$ Ha. $\Delta_{\mathrm{opt}}$ is reported in mHa.
}
\label{tab:si-diag-cr2}
\begin{tabular}{l r r r r}
\toprule
Active Space & $|\mathcal{V}|$ & $E_{\mathrm{opt}}$ / Ha & $E_{\mathrm{diag}}$ / Ha & $\Delta_{\mathrm{opt}}$ / mHa \\
\midrule
\multirow{5}{*}{CAS(24e, 30o)}
& 8 192     & -0.339793 & -0.339794 & 0.000 \\
& 32 768    & -0.372492 & -0.372499 & 0.006 \\
& 131 072   & -0.392029 & -0.392188 & 0.159 \\
& 524 288   & -0.402847 & -0.403853 & 1.006 \\
& 1 048 576 & -0.406173 & -0.407888 & 1.715 \\
\midrule
\multirow{5}{*}{CAS(48e, 42o)}
& 8 192     & -0.330898 & -0.330971 & 0.073 \\
& 32 768    & -0.381646 & -0.381673 & 0.028 \\
& 131 072   & -0.405377 & -0.405707 & 0.330 \\
& 524 288   & -0.418717 & -0.419859 & 1.142 \\
& 1 048 576 & -0.422971 & -0.424901 & 1.930 \\
\bottomrule
\end{tabular}
\end{table}

\clearpage

\section{Detailed Numerical Data}
\label{sec:si-data}

The following tables provide the numerical data used to generate the dissociation and timing figures in the main text.

\begin{longtable}{l S[table-format=-2.6] S[table-format=-2.6]}
    \caption{Detailed ground-state energies for the symmetric dissociation of \ce{H2O} in the cc-pVDZ basis set. $|\mathcal{V}|$ denotes the size of the variational subspace (Top-$K$). $E_{\text{var}}$ is the NQS variational energy, and $E_{\text{total}}$ includes the second-order Epstein-Nesbet perturbative correction ($E_{\text{var}} + \Delta E_{\text{PT2}}$). All values are reported in Hartrees ($E_h$).} \label{tab:h2o_pes_detailed} \\
    
    \toprule
    {Subspace size} & {$E_{\text{var}}$ ($E_h$)} & {$E_{\text{total}}$ ($E_h$)} \\
    \midrule
    \endfirsthead
    
    \multicolumn{3}{c}{Table \thetable{} (Continued)} \\
    \toprule
    {Subspace size} & {$E_{\text{var}}$ ($E_h$)} & {$E_{\text{total}}$ ($E_h$)} \\
    \midrule
    \endhead
    
    \multicolumn{3}{r}{Continued on next page...} \\
    \bottomrule
    \endfoot
    
    \bottomrule
    \endlastfoot

    %% --- 1.0 Re ---
    \multicolumn{3}{l}{$R = 1.0 R_e$} \\*
    \multicolumn{3}{l}{O(0.000000, 0.000000, 0.000000); H($\pm$1.010000, 0.758000, 0.000000)} \\*
    \midrule
    $|\mathcal{V}| = 1024$   & -76.207599 & -76.241762 \\
    $|\mathcal{V}| = 2048$   & -76.228652 & -76.241878 \\
    $|\mathcal{V}| = 4096$   & -76.233468 & -76.241836 \\
    $|\mathcal{V}| = 8192$   & -76.235706 & -76.241809 \\
    $|\mathcal{V}| = 16384$  & -76.237406 & -76.241805 \\
    $|\mathcal{V}| = 32768$  & -76.238839 & -76.241819 \\
    $|\mathcal{V}| = 65536$  & -76.239927 & -76.241831 \\
    $|\mathcal{V}| = 131072$ & -76.240633 & -76.241834 \\
    \addlinespace
    Benchmarks: & {UHF: -76.024039} & {MRCISD: -76.240434} \\*
                & {UCCSDT: -76.241367} & {FCI: -76.241860} \\*
    \midrule

    %% --- 1.5 Re ---
    \multicolumn{3}{l}{$R = 1.5 R_e$} \\
    \multicolumn{3}{l}{O(0.000000, 0.000000, -0.009000); H(0.000000, $\pm$1.515263, -1.058898)} \\
    \midrule
    $|\mathcal{V}| = 1024$   & -76.021217 & -76.071316 \\
    $|\mathcal{V}| = 2048$   & -76.049151 & -76.071181 \\
    $|\mathcal{V}| = 4096$   & -76.058468 & -76.071696 \\
    $|\mathcal{V}| = 8192$   & -76.063649 & -76.072036 \\
    $|\mathcal{V}| = 16384$  & -76.066736 & -76.072174 \\
    $|\mathcal{V}| = 32768$  & -76.068747 & -76.072260 \\
    $|\mathcal{V}| = 65536$  & -76.069903 & -76.072271 \\
    $|\mathcal{V}| = 131072$ & -76.070621 & -76.072275 \\
    \addlinespace
    Benchmarks: & {UHF: -75.829813} & {MRCISD: -76.071151} \\*
                & {UCCSDT: -76.070281} & {FCI: -76.072348} \\*
    \midrule

    %% --- 2.0 Re ---
    \multicolumn{3}{l}{$R = 2.0 R_e$} \\*
    \multicolumn{3}{l}{O(0.000000, 0.000000, -0.013500); H(0.000000, $\pm$2.272895, -1.588347)} \\*
    \midrule
    $|\mathcal{V}| = 1024$   & -75.905673 & -75.949154 \\
    $|\mathcal{V}| = 2048$   & -75.925819 & -75.950111 \\
    $|\mathcal{V}| = 4096$   & -75.938465 & -75.950860 \\
    $|\mathcal{V}| = 8192$   & -75.943786 & -75.951242 \\
    $|\mathcal{V}| = 16384$  & -75.947021 & -75.951467 \\
    $|\mathcal{V}| = 32768$  & -75.948855 & -75.951566 \\
    $|\mathcal{V}| = 65536$  & -75.949988 & -75.951610 \\
    $|\mathcal{V}| = 131072$ & -75.950464 & -75.951622 \\
    \addlinespace
    Benchmarks: & {UHF: -75.793668} & {MRCISD: -75.950722} \\*
                & {UCCSDT: -75.946153} & {FCI: -75.951665} \\*
    \midrule

    %% --- 2.5 Re ---
    \multicolumn{3}{l}{$R = 2.5 R_e$} \\*
    \multicolumn{3}{l}{O(0.000000, 0.000000, -0.018000); H(0.000000, $\pm$3.030526, -2.117796)} \\*
    \midrule
    $|\mathcal{V}| = 1024$   & -75.873290 & -75.914477 \\
    $|\mathcal{V}| = 2048$   & -75.899047 & -75.915713 \\
    $|\mathcal{V}| = 4096$   & -75.907279 & -75.916919 \\
    $|\mathcal{V}| = 8192$   & -75.911828 & -75.917529 \\
    $|\mathcal{V}| = 16384$  & -75.914525 & -75.917758 \\
    $|\mathcal{V}| = 32768$  & -75.916052 & -75.917851 \\
    $|\mathcal{V}| = 65536$  & -75.916768 & -75.917909 \\
    $|\mathcal{V}| = 131072$ & -75.917140 & -75.917931 \\
    \addlinespace
    Benchmarks: & {UHF: -75.791198} & {MRCISD: -75.917147} \\*
                & {UCCSDT: -75.916217} & {FCI: -75.917991} \\*
    \midrule

    %% --- 3.0 Re ---
    \multicolumn{3}{l}{$R = 3.0 R_e$} \\*
    \multicolumn{3}{l}{O(0.000000, 0.000000, -0.022500); H(0.000000, $\pm$3.788158, -2.647245)} \\*
    \midrule
    $|\mathcal{V}| = 1024$   & -75.905255 & -75.910913 \\
    $|\mathcal{V}| = 2048$   & -75.898731 & -75.909550 \\
    $|\mathcal{V}| = 4096$   & -75.909467 & -75.911011 \\
    $|\mathcal{V}| = 8192$   & -75.910387 & -75.911030 \\
    $|\mathcal{V}| = 16384$  & -75.910780 & -75.911060 \\
    $|\mathcal{V}| = 32768$  & -75.910915 & -75.911088 \\
    $|\mathcal{V}| = 65536$  & -75.910986 & -75.911130 \\
    $|\mathcal{V}| = 131072$ & -75.911020 & -75.911144 \\
    \addlinespace
    Benchmarks: & {UHF: -75.790955} & {MRCISD: -75.911121} \\*
                & {UCCSDT: -75.911407} & {FCI: -75.911946} \\*
\end{longtable}
\begin{longtable}{l S[table-format=-3.6] S[table-format=-3.6]}
    \caption{Detailed ground-state energies for the bond breaking of \ce{N2} in the cc-pVDZ basis set. $|\mathcal{V}|$ denotes the size of the variational subspace (Top-$K$). $E_{\text{var}}$ is the NQS variational energy, and $E_{\text{total}}$ includes the second-order Epstein-Nesbet perturbative correction ($E_{\text{var}} + \Delta E_{\text{PT2}}$). All values are reported in Hartrees ($E_h$).} \label{tab:n2_pes_detailed} \\
    
    \toprule
    {Subspace size} & {$E_{\text{var}}$ ($E_h$)} & {$E_{\text{total}}$ ($E_h$)} \\
    \midrule
    \endfirsthead
    
    \multicolumn{3}{c}{Table \thetable{} (Continued)} \\
    \toprule
    {Subspace size} & {$E_{\text{var}}$ ($E_h$)} & {$E_{\text{total}}$ ($E_h$)} \\
    \midrule
    \endhead
    
    \midrule
    \multicolumn{3}{r}{Continued on next page...} \\
    \bottomrule
    \endfoot
    
    \bottomrule
    \endlastfoot

    %% --- 2.118 Bohr ---
    \multicolumn{3}{l}{$d = 2.118$ Bohr} \\*
    \midrule
    $|\mathcal{V}| = 1024$   & -109.195580 & -109.281924 \\
    $|\mathcal{V}| = 2048$   & -109.232688 & -109.281728 \\
    $|\mathcal{V}| = 4096$   & -109.252369 & -109.282155 \\
    $|\mathcal{V}| = 8192$   & -109.260943 & -109.282143 \\
    $|\mathcal{V}| = 16384$  & -109.266361 & -109.282030 \\
    $|\mathcal{V}| = 32768$  & -109.270646 & -109.281990 \\
    $|\mathcal{V}| = 65536$  & -109.273813 & -109.281976 \\
    $|\mathcal{V}| = 131072$ & -109.276065 & -109.282082 \\
    $|\mathcal{V}| = 262144$ & -109.277462 & -109.282036 \\
    $|\mathcal{V}| = 524288$ & -109.278217 & -109.282096 \\
    \addlinespace
    Benchmarks: & {UHF: -108.949378} & {MRCISD: -109.275356} \\
                & {UCCSDT: -109.280323} & {DMRG: -109.282157} \\
                & {CDFCI: -109.282173} & \\
    \midrule

    %% --- 2.4 Bohr ---
    \multicolumn{3}{l}{$d = 2.4$ Bohr} \\*
    \midrule
    $|\mathcal{V}| = 1024$   & -109.152881 & -109.240336 \\
    $|\mathcal{V}| = 2048$   & -109.188582 & -109.240568 \\
    $|\mathcal{V}| = 4096$   & -109.209565 & -109.241353 \\
    $|\mathcal{V}| = 8192$   & -109.218955 & -109.241157 \\
    $|\mathcal{V}| = 16384$  & -109.225309 & -109.241461 \\
    $|\mathcal{V}| = 32768$  & -109.229865 & -109.241542 \\
    $|\mathcal{V}| = 65536$  & -109.233145 & -109.241726 \\
    $|\mathcal{V}| = 131072$ & -109.235447 & -109.241712 \\
    $|\mathcal{V}| = 262144$ & -109.236950 & -109.241755 \\
    $|\mathcal{V}| = 524288$ & -109.237776 & -109.241755 \\
    \addlinespace
    Benchmarks: & {UHF: -108.891623} & {MRCISD: -109.234925} \\
                & {UCCSDT: -109.238030} & {DMRG: -109.241886} \\
                & {CDFCI: -109.241908} & \\
    \midrule

    %% --- 2.7 Bohr ---
    \multicolumn{3}{l}{$d = 2.7$ Bohr} \\*
    \midrule
    $|\mathcal{V}| = 1024$   & -109.060701 & -109.161485 \\
    $|\mathcal{V}| = 2048$   & -109.096580 & -109.161369 \\
    $|\mathcal{V}| = 4096$   & -109.122858 & -109.162392 \\
    $|\mathcal{V}| = 8192$   & -109.134548 & -109.162687 \\
    $|\mathcal{V}| = 16384$  & -109.142673 & -109.162677 \\
    $|\mathcal{V}| = 32768$  & -109.148413 & -109.162932 \\
    $|\mathcal{V}| = 65536$  & -109.152122 & -109.162961 \\
    $|\mathcal{V}| = 131072$ & -109.154866 & -109.163218 \\
    $|\mathcal{V}| = 262144$ & -109.156426 & -109.163357 \\
    $|\mathcal{V}| = 524288$ & -109.157566 & -109.163305 \\
    \addlinespace
    Benchmarks: & {UHF: -108.833687} & {MRCISD: -109.156473} \\
                & {UCCSDT: -109.156703} & {DMRG: -109.163572} \\
                & {CDFCI: -109.163600} & \\
    \midrule

    %% --- 3.0 Bohr ---
    \multicolumn{3}{l}{$d = 3.0$ Bohr} \\*
    \midrule
    $|\mathcal{V}| = 1024$   & -108.971587 & -109.084199 \\
    $|\mathcal{V}| = 2048$   & -109.015137 & -109.084913 \\
    $|\mathcal{V}| = 4096$   & -109.041489 & -109.086501 \\
    $|\mathcal{V}| = 8192$   & -109.056459 & -109.087600 \\
    $|\mathcal{V}| = 16384$  & -109.066061 & -109.087913 \\
    $|\mathcal{V}| = 32768$  & -109.072945 & -109.088328 \\
    $|\mathcal{V}| = 65536$  & -109.077329 & -109.089105 \\
    $|\mathcal{V}| = 131072$ & -109.080326 & -109.088924 \\
    $|\mathcal{V}| = 262144$ & -109.082139 & -109.088994 \\
    $|\mathcal{V}| = 524288$ & -109.083133 & -109.089032 \\
    \addlinespace
    Benchmarks: & {UHF: -108.790272} & {MRCISD: -109.082149} \\
                & {UCCSDT: -109.079437} & {DMRG: -109.089380} \\
                & {CDFCI: -109.089405} & \\
    \midrule

    %% --- 3.6 Bohr ---
    \multicolumn{3}{l}{$d = 3.6$ Bohr} \\*
    \midrule
    $|\mathcal{V}| = 1024$   & -108.854881 & -108.983581 \\
    $|\mathcal{V}| = 2048$   & -108.888278 & -108.985648 \\
    $|\mathcal{V}| = 4096$   & -108.922916 & -108.988277 \\
    $|\mathcal{V}| = 8192$   & -108.948063 & -108.991807 \\
    $|\mathcal{V}| = 16384$  & -108.964993 & -108.994206 \\
    $|\mathcal{V}| = 32768$  & -108.976022 & -108.995841 \\
    $|\mathcal{V}| = 65536$  & -108.982483 & -108.996639 \\
    $|\mathcal{V}| = 131072$ & -108.985929 & -108.997008 \\
    $|\mathcal{V}| = 262144$ & -108.987791 & -108.997119 \\
    $|\mathcal{V}| = 524288$ & -108.988951 & -108.997068 \\
    \addlinespace
    Benchmarks: & {UHF: -108.767549} & {MRCISD: -108.990759} \\
                & {UCCSDT: -108.990518} & {DMRG: -108.998052} \\
                & {CDFCI: -108.998083} & \\
    \midrule

    %% --- 4.2 Bohr ---
    \multicolumn{3}{l}{$d = 4.2$ Bohr} \\*
    \midrule
    $|\mathcal{V}| = 1024$   & -108.818182 & -108.944537 \\
    $|\mathcal{V}| = 2048$   & -108.859307 & -108.949226 \\
    $|\mathcal{V}| = 4096$   & -108.893786 & -108.955348 \\
    $|\mathcal{V}| = 8192$   & -108.925608 & -108.963166 \\
    $|\mathcal{V}| = 16384$  & -108.943700 & -108.965669 \\
    $|\mathcal{V}| = 32768$  & -108.951043 & -108.966821 \\
    $|\mathcal{V}| = 65536$  & -108.955324 & -108.967338 \\
    $|\mathcal{V}| = 131072$ & -108.958120 & -108.967493 \\
    $|\mathcal{V}| = 262144$ & -108.959330 & -108.967648 \\
    $|\mathcal{V}| = 524288$ & -108.960053 & -108.967866 \\
    \addlinespace
    Benchmarks: & {UHF: -108.775057} & {MRCISD: -108.963070} \\
                & {UCCSDT: -108.966852} & {DMRG: -108.970090} \\
                & {CDFCI: -108.970132} & \\
\end{longtable}
\begin{longtable}{l r S[table-format=5.2] S[table-format=5.2] S[table-format=5.2] S[table-format=5.2]}
    \caption{Detailed wall-time breakdown of 30 outer-loop iterations for various systems. $|\mathcal{V}|$ denotes the size of the variational subspace (Top-$K$). All times are reported in seconds (s).} \label{tab:timing_breakdown_si} \\
    
    \toprule
    {System} & {$|\mathcal{V}|$} & {Compilation} & {Inner} & {Overhead} & {Total Wall-time} \\
    \midrule
    \endfirsthead
    
    \multicolumn{6}{c}{Table \thetable{} (Continued)} \\
    \toprule
    {System} & {$|\mathcal{V}|$} & {Compilation} & {Inner} & {Overhead} & {Total Wall-time} \\
    \midrule
    \endhead
    
    \multicolumn{6}{r}{Continued on next page...} \\
    \bottomrule
    \endfoot
    
    \bottomrule
    \endlastfoot

    %% --- H2O ---
    \ce{H2O} 1.0$R_e$ & 2048 & 12.24 & 82.37 & 44.89 & 140.98 \\*
    (cc-pVDZ) & 8192 & 40.86 & 204.28 & 46.45 & 293.10 \\*
    & 32768 & 153.38 & 806.72 & 66.23 & 1027.91 \\*
    & 131072 & 601.36 & 3376.98 & 89.91 & 4070.34 \\*
    \midrule

    %% --- N2 ---
    \ce{N2} 2.118 Bohr & 2048 & 28.18 & 83.82 & 77.01 & 190.93 \\
    (cc-pVDZ) & 8192 & 96.44 & 228.64 & 128.01 & 455.01 \\
    & 32768 & 346.50 & 834.70 & 231.87 & 1415.08 \\
    & 131072 & 1224.77 & 3020.84 & 363.99 & 4612.43 \\
    \midrule

    %% --- Cr2 24e ---
    \ce{Cr2} & 8192 & 166.40 & 267.19 & 307.48 & 744.18 \\
    (24e, 30o) & 32768 & 614.99 & 1047.20 & 721.66 & 2387.10 \\
    & 131072 & 2312.76 & 3996.06 & 1463.95 & 7776.67 \\
    & 524288 & 9608.18 & 17684.18 & 2245.62 & 29544.14 \\
    \midrule

    %% --- Cr2 48e ---
    \ce{Cr2} & 8192 & 739.81 & 541.39 & 2440.75 & 3728.74 \\
    (48e, 42o) & 32768 & 2680.34 & 2345.80 & 5794.15 & 10827.40 \\
    & 131072 & 9711.81 & 9091.87 & 11468.47 & 30279.84 \\
    & 524288 & 32118.66 & 38870.37 & 15284.47 & 86283.72 \\

\end{longtable}

\end{document}